\documentclass[12pt,a4paper,hypertex]{article}
\usepackage{jheppub_modified}

\allowdisplaybreaks[1]
\usepackage{amsmath,bm}
\usepackage{subfigure}
\usepackage{amssymb}
\title{
Photonic black resonators and photon stars in AdS$_5$
}

\author[1]{Takaaki Ishii,}
\author[2]{Keiju Murata}

\affiliation[1]{Department of Physics, Kyoto University, Kyoto 606-8502, Japan}
\affiliation[2]{Department of Physics, College of Humanities and Sciences, Nihon University, Tokyo 156-8550, Japan}
\emailAdd{ishiitk@gauge.scphys.kyoto-u.ac.jp, murata.keiju@nihon-u.ac.jp}

\abstract{%
Rapidly rotating Myers-Perry-AdS$_5$ (MPAdS$_5$) black holes are shown to be unstable against rotational superradiance of a Maxwell field. From the onset of the instability, time-periodic neutral black hole solutions equipped with a nontrivial electromagnetic wave are obtained, which we call {\it photonic black resonators}. In the horizonless limit, they reduce to geon solutions which may be called {\it photon stars}. Specifically, we introduce a cohomogeneity-1 ansatz for the metric and Maxwell field and construct such solutions with an $R\times SU(2)$ isometry group. We compute thermodynamic quantities and obtain phase diagrams. It turns out that a photonic black resonator has a higher entropy than a MPAdS$_5$ black hole, while it also has a smaller entropy than a black resonator without the Maxwell field. This suggests what is expected for nonlinear dynamics following the Maxwell superradiant instability with the $SU(2)$ isometry.}

\preprint{KUNS-2773}

\begin{document}
\maketitle

\section{Introduction}

A variety of black hole solutions have been investigated in higher dimensions and asymptotically anti de Sitter (AdS) spacetime.
In higher dimensions, extra spatial directions accommodate more symmetries, and various types of black hole solutions can be constructed (see \cite{Emparan:2008eg} for a review).
The construction of black hole solutions 
in asymptotically AdS spacetime has been motivated to study in line with the development of the AdS/CFT duality \cite{Maldacena:1997re,Gubser:1998bc,Witten:1998qj}.

We focus on rotating black holes in five-dimensional asymptotically AdS spacetime (AdS$_5$).
The Myers-Perry black hole~\cite{Myers:1986un}, which is the higher dimensional generalization of the Kerr black hole solution, was generalized to include a cosmological constant in \cite{Hawking:1998kw,Gibbons:2004uw,Gibbons:2004js}.
The thermodynamics of Myers-Perry-AdS (MPAdS) black holes was closely discussed in \cite{Gibbons:2004ai} (see also \cite{Papadimitriou:2005ii}). There are phase transitions for the rotating black hole solutions \cite{Carter:2005uw}.

Superradiant instability is one of the most drastic phenomena that rotating black holes exhibit in AdS spacetime \cite{Hawking:1999dp,Cardoso:2004hs}.
This is caused by the repetition of the wave amplification by superradiance and the reflection of the wave by the boundary of the AdS spacetime.
In Kerr-AdS$_4$ and MPAdS spacetimes, 
the superradiant instability has been studied for gravitational perturbations in~\cite{Cardoso:2006wa,Kunduri:2006qa,Murata:2008xr,Kodama:2009rq,Cardoso:2013pza}. (See also \cite{Brito:2015oca} for a review.) 
At the onset of the superradiant instability, there is a normal mode.

By the nonlinear extension of the normal mode, a new family of dynamical black hole solutions is expected to branch from the rotating black holes at the onset of the gravitational superradiant instability~\cite{Kunduri:2006qa}. 
Such black holes have been first constructed in Ref.~\cite{Dias:2015rxy} and named {\it black resonators} because these represent time periodic black holes, having less isometries than the original stationary black holes.
In particular, we will call such black resonator solutions obtained in pure Einstein gravity as {\it gravitational black resonators}
in order to distinguish them from the solutions we will construct in the presence of a Maxwell field in this paper.
In AdS$_4$, the construction of the gravitational black resonators was done by solving three-dimensional partial differential equations without assuming spatial isometries~\cite{Dias:2015rxy}. In AdS$_5$, in contrast, a cohomogeneity-1 metric ansatz for gravitational black resonators with an $R\times SU(2)$ isometry group has been proposed, and solutions were obtained by solving ordinary differential equations (ODEs)~\cite{Ishii:2018oms}. The horizonless limits of black resonators are geons, which can be obtained as the nonlinear extension of the normal modes of AdS~\cite{Dias:2011ss,Horowitz:2014hja,Martinon:2017uyo,Fodor:2017spc,Ishii:2018oms}. 
In previous works, black resonators and geons have been considered in pure Einstein gravity.

In this paper, we consider time periodic spacetimes induced by a dynamical Maxwell field in AdS$_5$. The stability against Maxwell perturbations has been proven in Ref.~\cite{Hawking:1999dp} for the Kerr-AdS$_4$ and MPAdS with the angular velocities $\Omega<1$ in units of the AdS radius. In $\Omega>1$, however, Maxwell perturbations also can cause the superradiant instability.\footnote{In this paper, we consider only the rotational superradiant instability and not the charged one. Therefore, we may simply use the term {\it superradiant instability} to represent the rotational one.} For Kerr-AdS$_4$, the Maxwell superradiant instability is shown in Ref.~\cite{Wang:2015fgp}. In MPAdS$_5$, we will study it in section~\ref{Maxwell}.
From the onset of the Maxwell rotational superradiant instability, we can extend a new family of time periodic black hole solutions. We will call such solutions equipped with an electromagnetic wave as {\it photonic black resonators}. Their horizonless limit may be called {\it photon stars}. Based on our previous work~\cite{Ishii:2018oms}, we introduce a cohomogeneity-1 ansatz for the metric and Maxwell field with an $R\times SU(2)$ isometry group. With this ansatz, the Einstein and Maxwell equations reduce to ODEs. Solving them, we construct photonic black resonators and photon stars.

This paper is organized as follows.
In section~\ref{Maxwell}, we study the Maxwell perturbation in MPAdS$_5$.
We evaluate the onset of the superradiant instability for $SU(2)$-symmetric perturbations.
In section~\ref{coh1}, we introduce the cohomogeneity-1 ansatz for metric and the Maxwell field for 
photonic black resonators and photon stars. We also write down the complete set of Einstein and Maxwell equations given by ODEs. 
In section~\ref{phstr}, 
we construct photon star solutions by perturbative and fully numerical methods.
In section~\ref{phtoBR}, we obtain photonic black resonators numerically
and compute their thermodynamical quantities.
We also discuss the phase diagram of the photonic black resonators and photon stars.
We conclude in section~\ref{conc} with a summary and discussion.

\section{Maxwell superradiant instability of Myers-Perry-AdS$_5$ with equal angular momenta}
\label{Maxwell}

In this section, we study a Maxwell perturbation of MPAdS$_5$
with equal angular momenta.
We first introduce the setup for MPAdS$_5$ and then evaluate the superradiant instability by a Maxwell field.

\subsection{Setup}
We consider the Einstein-Maxwell theory in asymptotically global AdS$_5$ spacetime. 
The action is given by\footnote{
We do not consider the Chern-Simons term in this paper because 
it breaks 
parity symmetry and makes the ansatz for the metric and Maxwell field complicated. (See section~\ref{C1anzatz}.)
Therefore, photonic black resonators and photon stars which will be constructed in this paper are not 
solutions in the five-dimensional minimal gauged supergravity. 
However, we could generalize our solutions 
as those in $U(1)^3$-gauged $\mathcal{N}=2$ five-dimensional supergravity, which has 
three $U(1)$ gauge fields $A^1,A^2,A^3$. 
If we set $A^1=A^2=0$ and $A^3=A$, 
the Chern-Simons term does not contribute to the equations of motion.
Then, we would be able to construct photonic black resonators and photon stars in the $U(1)^3$-gauged supergravity by using our metric ansatz, while we will also need to consider other fields in the theory.
We would like to postpone this direction as a future work.
}
\begin{equation}
S=\frac{1}{16\pi G_5}\int d^5 x \sqrt{-g} \left[R+\frac{12}{L^2}-\frac{1}{4} F_{\mu\nu} F^{\mu\nu}\right]\ ,
\label{action}
\end{equation}
where $G_5$ is the five-dimensional Newton's constant and $L$ is the AdS radius.
The normalization of the gauge field $A_\mu$ is chosen so that the Newton's constant is factored out in the action.

The Myers-Perry-AdS black hole is an exact solution of (\ref{action}).
Throughout this paper we consider only uncharged black holes, and we set the gauge field trivial for the time being: $A_\mu=0$.
While in general the form of the Myers-Perry black hole solution is cumbersome, in the case of equal angular momenta the solution is cohomogeneity-1 and the metric can be simply written as
\begin{multline}
 ds^2=-(1+r^2)f(r)d\tau^2 + \frac{dr^2}{(1+r^2)g(r)}\\+\frac{r^2}{4} \left[
\sigma_1^2 + \sigma_2^2 + \beta(r)(\sigma_3+2 h(r)d\tau)^2
\right]\ ,
\label{MPAdSmetric}
\end{multline}
where we introduced $SU(2)$-invariant 1-forms $\sigma_{1,2,3}$ as
\begin{equation}
\begin{split}
\sigma_1 &= -\sin\chi d\theta + \cos\chi\sin\theta d\phi\ ,\\
\sigma_2 &= \cos\chi d\theta + \sin\chi\sin\theta d\phi\ ,\\
\sigma_3 &= d\chi + \cos\theta d\phi  \ . 
\end{split}
\label{invf}
\end{equation}
Here, $(\theta,\phi,\chi)$ are angular coordinates on $S^3$. Their ranges are $0\leq \theta \leq \pi $, $0\leq \phi <2\pi$, and $0\leq \chi <4\pi$ with the periodicity  of a twisted torus $(\theta,\phi,\chi) \sim (\theta,\phi+2\pi,\chi+2\pi) \sim (\theta,\phi,\chi+4\pi)$.
The functions in the components of the metric are explicitly given by
\begin{equation}
\begin{split}
g(r)&=1-\frac{2\mu (1-a^2)}{r^2(1+r^2)} +\frac{2a^2\mu}{r^4(1+r^2)}\ ,\quad
\beta(r)=1+\frac{2 a^2\mu}{r^4}\ ,\\
h(r)&=\Omega-\frac{2\mu a}{r^4+2 a^2\mu}\ ,\quad 
f(r)=\frac{g(r)}{\beta(r)}\ ,
\end{split}
\label{MPAdSFunctions}
\end{equation}
where $\mu, a, \Omega$ are parameters.
The event horizon is located at $r=r_h$ defined by the largest root of $g(r)$.
In $h(r)$, there is a freedom to choose the constant $\Omega$.
We fix it so that $h(r_h)=0$:
\begin{equation}
\Omega=\frac{2\mu a}{r_h^4+2 a^2\mu}\ .
\label{MPAdSOmega}
\end{equation}
Then, $\partial_\tau$ becomes the horizon Killing vector: $g_{\tau\tau}|_{r=r_h}=0$.
We will consider only MPAdS$_5$ with equal angular momenta hereafter.

The isometry group of the equal angular momentum solution is $R \times U(2)$ where $R$ denotes $\tau$-translations generated by $\partial_\tau$.
That the dependence on the angular coordinates $(\theta,\psi,\chi)$ is only through the $SU(2)$-invariant 1-forms indicates that the metric (\ref{MPAdSmetric}) possesses an $SU(2)$-isometry.
This spacetime also has a $U(1)$-isometry generated by $\partial_\chi$.
This rotates $\sigma_1$ and $\sigma_2$:
Introducing $\sigma_\pm$ as
\begin{equation}
 \sigma_\pm \equiv \frac{1}{2}(\sigma_1 \mp i\sigma_2)\ ,
\end{equation}
we obtain 
\begin{equation}
 \mathcal{L}_{i\partial_\chi}\sigma_\pm = \pm \sigma_\pm\ ,
\end{equation}
where $\mathcal{L}$ denotes the Lie derivative.
Thus the 1-forms $\sigma_\pm$ have $U(1)$-charges $\pm 1$, respectively.
However, the combination $\sigma_1^2+\sigma_2^2=4\sigma_+ \sigma_-$ is neutral under the $U(1)$-rotation, 
and the spacetime~(\ref{MPAdSmetric}) is invariant under $\partial_\chi$ although each $\sigma_{1,2}$ is rotated.
In summary, the isometry group of the spacetime~(\ref{MPAdSmetric}) is 
$R\times SU(2)\times U(1) \simeq R \times U(2)$.

The choice of $h(r)$ employed in (\ref{MPAdSFunctions}) with (\ref{MPAdSOmega}) corresponds to the {\it rotating frame at infinity}.
We have $h(r_h)=0$ and $h(\infty)=\Omega$, and this indicates that the AdS boundary, locating at $r=\infty$, is rotating in this coordinate frame.
On the other hand, one can also go to the {\it non-rotating frame at infinity} by
\begin{equation}
dt=d\tau\ ,\quad d\psi=d\chi+2 \Omega d\tau\ .
\label{psishift}
\end{equation}
In the new frame, we introduce $\bar{h}(r) \equiv h(r) - \Omega$ together with the 1-forms in that frame $\bar{\sigma}_a$ as $\bar{\sigma}_1 =-\sin\psi d\theta + \cos\psi\sin\theta d\phi$, etc.
With the bar-notations, the metric (\ref{MPAdSmetric}) remains the same form under the frame change.
The behavior of $\bar{h}(r)$ is $\bar{h}(r_h)=-\Omega$ and $\bar{h}(\infty)=0$, which means that the AdS boundary is not rotating.
In the non-rotating frame at infinity, the horizon Killing vector is helical: $\partial_\tau = \partial_t + \Omega \partial_{\psi/2}$.\footnote{We define $\Omega$ so that it is the angular velocity with respect to $\psi/2 \in [0,2\pi)$. This definition matches other literature such as~\cite{Gibbons:2004ai,Kunduri:2006qa,Murata:2008xr}.}

\subsection{Onset of Maxwell superradiant instability}

Let us consider a Maxwell perturbation which preserves the $SU(2)$ isometry in the spacetime~(\ref{MPAdSmetric}). 
To discuss the onset of a superradiant instability, it is sufficient to use a $\tau$-independent perturbation because a normal mode is induced at the onset of the instability.
Then our task is to identify the value of $\Omega$ when the normal mode arises.
We introduce an $SU(2)$-invariant perturbation given by
\begin{equation}
 A=\gamma(r)\sigma_1=\gamma(r)(\sigma_+ + \sigma_-)\ .
\label{Mxwpert}
\end{equation}
This has $U(1)$-charges $\pm 1$ and hence breaks the $U(1)$ isometry. 
The other components of $SU(2)$-invariant Maxwell perturbations are all neutral under the $U(1)$-rotation.
(Namely, $d\tau$, $dr$ and $\sigma_3$ are invariant under $\partial_\chi$.)
Therefore, the perturbation (\ref{Mxwpert}) decouples from the other Maxwell perturbations.
For (\ref{Mxwpert}), the Maxwell equation $\nabla^\mu F_{\mu\nu}=0$ gives
\begin{equation}
 \gamma''=-\bigg\{
\frac{g'}{g}
+\frac{1+3r^2}{r(1+r^2)}\bigg\}\gamma'
-\frac{4}{(1+r^2)fg^2}\bigg\{\frac{gh^2}{{1+r^2}}  -\frac{f^2}{r^2}\bigg\}\gamma\ .
\label{pbr_onset_eq}
\end{equation}

If the background is pure AdS, i.e.~$f=g=1$, the onset can be analytically obtained.
The perturbation equation \eqref{pbr_onset_eq} can be analytically solved by
\begin{equation}
\gamma(r) = r^2(1+r^2)^{-\Omega} {}_2F_1(1-\Omega,2-\Omega;3;-r^2) \ ,
\label{linear_gamma_sol}
\end{equation}
where regularity was imposed at the center of the AdS, $r=0$.
Near the AdS boundary $r \to \infty$, the solution behaves as
\begin{equation}
\gamma(r) = \frac{2}{\Gamma(2-\Omega) \Gamma(2+\Omega)} + \cdots \ .
\end{equation}
The Dirichlet boundary condition $\gamma(r)|_{r=\infty} = 0$ is satisfied only if $\Omega = 2+n \ (n=0,1,2,\cdots)$.
We hence obtain a tower of normal modes in global AdS$_5$.

In the MPAdS$_5$ background, we solve \eqref{pbr_onset_eq} numerically with the Dirichlet boundary condition at $r=\infty$ as well as regularity at $r=r_h$.
Near the horizon, the regular solution approaches a constant. 
Imposing this behavior, we integrate (\ref{pbr_onset_eq}) from $r=r_h$ to $r=\infty$ and find the values of ($r_h,\Omega$) so that $\gamma=0$ is satisfied at $r=\infty$.
Such ($r_h,\Omega$) correspond to the onset of the superradiant instability.
Depending on the number of nodes between $r=r_h$ and $r=\infty$, we obtain a tower of normal modes.
We label them by $n=0,1,2,\cdots$ and call $n=0$ the fundamental tone and $n \ge 1$ overtones.

\begin{figure}[t]
\centering
\includegraphics[scale=0.5]{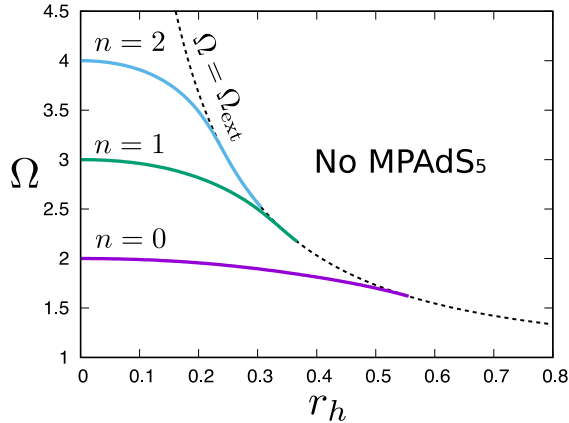}
\caption{%
Onset for the Maxwell superradiant instability. The angular velocity of the extreme MPAdS$_5$ black hole $\Omega=\Omega_\mathrm{ext}$ is plotted in the dashed line.
There are no regular MPAdS$_5$ solutions in the upper right region.}
\label{onset}
\end{figure}

In Fig.~\ref{onset}, the instability frequencies are shown for the $n=0,1,2$ modes.
It is expected that the onset curves collide the frequency of the extreme MPAdS$_5$ black hole.
In appendix~\ref{app:ext}, we directly evaluate the linear perturbation on top of the extreme black hole and find that the endpoints of the onset curves for the $n=0$ and $n=1$ modes would be located at $r_h = 0.5559$ and  $0.3720$, respectively.

\section{Cohomogeneity-1 photonic black resonators}
\label{coh1}

In this section, we introduce the ansatz for cohomogeneity-1 black resonators equipped with a nontrivial Maxwell field.

\subsection{Ansatz for metric and Maxwell field}
\label{C1anzatz}

We will take into account the backreaction of the Maxwell perturbation~(\ref{Mxwpert}) 
to the metric, i.e., we will solve the full nonlinear equations of motion derived from the action~(\ref{action}).
For the Maxwell field, we assume the form given in (\ref{Mxwpert}).
For the metric, we use the cohomogeneity-1 metric ansatz introduced in Ref.~\cite{Ishii:2018oms}:
\begin{multline}
 ds^2=-(1+r^2)f(r)d\tau^2 + \frac{dr^2}{(1+r^2)g(r)}\\
+\frac{r^2}{4} \left[
\alpha(r)\sigma_1^2 + \frac{1}{\alpha(r)}\sigma_2^2 + \beta(r)(\sigma_3+2 h(r)d\tau)^2
\right]\ ,
\label{metricanz}
\end{multline}
where, compared with (\ref{MPAdSmetric}), we introduced an extra function $\alpha(r)$ which deforms the $S^2$ base space.
The isometry group preserved by the metric (\ref{metricanz}) is $R \times SU(2)$.

We can show that Eqs.~(\ref{Mxwpert}) and (\ref{metricanz}) form a consistent ansatz as follows.
If the $\tau$-translation and $SU(2)$ symmetries are assumed, 
we can write the Maxwell field and metric as  
$A=A_a(r)e^a$ and $ds^2=g_{ab}(r)e^a e^b$ where $e^a=(dt,dr,\sigma_1,\sigma_2,\sigma_3)$.
Let us consider parity transformations $P_1$ and $P_2$ defined as
\begin{equation}
\begin{split}
P_1(\tau,\chi,\phi,\theta)&= (-\tau,-\chi,-\phi,\theta)\ ,\\
P_2(\tau,\chi,\phi,\theta)&= (-\tau,-\chi,\phi,\pi-\theta)\ .
\end{split}
\end{equation}
By these transformations, the 1-forms $e^a$ are transformed as 
\begin{equation}
\begin{split}
 P_1(d\tau,dr,\sigma_1,\sigma_2,\sigma_3)&= (-d\tau,dr,-\sigma_1,\sigma_2,-\sigma_3)\ ,\\
 P_2(d\tau,dr,\sigma_1,\sigma_2,\sigma_3)&= (-d\tau,dr,\sigma_1,-\sigma_2,-\sigma_3)\ .
\end{split}
\end{equation}
We choose the Maxwell perturbation to be even under $P_2$.
Then we have $A=A_r(r)dr+\gamma(r) \sigma_1$.\footnote{
Adopting $P_1$ instead corresponds to consider $A=A_r(r)dr+\gamma(r) \sigma_2$.
}
Using the $U(1)$ gauge freedom, we can impose $A_r=0$ and thus obtain (\ref{Mxwpert}).
We assume that the metric is even under both $P_1$ and $P_2$.
Then we obtain the metric~(\ref{metricanz}). 
While the Maxwell field~(\ref{Mxwpert}) is odd under the parity $P_1$,
the energy momentum tensor made out of it is quadratic in the gauge field and therefore is even under $P_1$.
Thus the Maxwell field (\ref{Mxwpert}) is consistent with the metric ansatz (\ref{metricanz}).\footnote{
This argument does not apply if the Chern-Simons term exists, which breaks the parity symmetry. 
We can still impose cohomogeneity-1 to the ansatz for the metric and Maxwell field even in the presence of the Chern-Simons term.
However, we need to use a much more complicated ansatz without assuming the parity symmetry. 
It would be interesting to construct such solutions in a future work.
}

Substituting the ansatz~(\ref{Mxwpert}) and (\ref{metricanz}) into the Einstein and Maxwell equations, 
we obtain the complete set of the equations of motion as
\begin{align}
\begin{split}
f'=&\frac{1}{r^3(1+r^2)^2 g \alpha^2 (r\beta'+6\beta)}
[
4 r^4 h^2 (\alpha^2-1)^2 \beta\\
&+r^3 (r^2+1) g\{r(1+r^2)  f \alpha'{}^2 \beta  
-r^5   h'{}^2 \alpha^2 \beta^2 
-2 r^2 (2+3 r^2)  f \alpha^2   \beta'\}\\
&
-4 r^2 (1+r^2) f  \{6 r^2 \alpha^2 \beta (g-1)+3 g \alpha^2 \beta + (\alpha^2-\alpha \beta+1)^2-4 \alpha^2\}\\
&
+4  r^2(1+r^2)^2f g\alpha\beta \gamma'{}^2
+16\alpha^3\{r^2\beta h^2-(1+r^2)f\}\gamma^2
]\ ,
\label{EOMf}
\end{split}
\\
\begin{split}
g'=&\frac{1}{6 r(1+r^2)^2 f \alpha^2 \beta}
[
-4 r^2 h^2   (\alpha^2-1)^2\beta\\
&+r(1+r^2)g\{
-r (1+r^2)  f \alpha'{}^2 \beta
+r^3 h'{}^2 \alpha^2 \beta^2\\
&- (-r(1+r^2) f'+2 f) \alpha^2 \beta'\}
+4 (1+r^2) f  \{-6r^2\alpha^2 \beta(g - 1)-3 g \alpha^2 \beta \\
& 
+\alpha^4+4 \alpha^3 \beta-5 \alpha^2 \beta^2-2 \alpha^2+4 \alpha \beta+1\}
-8 (1+r^2)^2fg\alpha\beta \gamma'{}^2
]\ ,
\label{EOMg}
\end{split}
\\
\begin{split}
h''=&\frac{1}{2 r^4 (1+r^2) f g \alpha^2 \beta }
[
8 r^2 f h (\alpha^2-1)^2\\
&-r^3 (1+r^2) h' \alpha^2 \{r (f g' \beta -f' g \beta +3 f g \beta')+10 f g \beta \}
+32f h \alpha^3 \gamma^2
]\ ,
\label{EOMh}
\end{split}
\\
\begin{split}
\alpha''=&\frac{1}{2 r^4 (1+r^2)^2 f  g \alpha \beta}
[
2 r^4 (r^2+1)^2  f g \alpha'{}^2 \beta\\
&-r^3 (r^2+1) \alpha \alpha'  \{r(1+r^2)(f g \beta)' +2 (3+5 r^2)  f g \beta\}\\
&-8 r^2 (\alpha^2-1) \{
r^2 h^2 \beta (\alpha^2 + 1)-(1+r^2) f \alpha(\alpha-\beta) -(1+r^2)f
\}\\
&
-4 r^2(1+r^2)^2fg \alpha\beta \gamma'{}^2
-16\alpha^3\{r^2h^2 \beta -(1+r^2)f\}\gamma^2
]\ ,
\label{EOMa}
\end{split}
\\
\begin{split}
\beta''=&\frac{1}{2 r^4 (1+r^2) f g \alpha^2 \beta}
[
-2r^6  g h'{}^2 \alpha^2 \beta^3 \\
&-r^3 \alpha^2 \beta' \{
r(1+r^2)(f' g \beta+ f g' \beta- f g \beta')
+2 (3+5 r^2)f g \beta \}\\
&-8 r^2 f \beta (\alpha^4+\alpha^3 \beta-2 \alpha^2 \beta^2-2 \alpha^2+\alpha \beta+1)\\
&
+4\ r^2(1+r^2)^2fg \alpha\beta^2\gamma'{}^2
-16\alpha^3\beta \{r^2 h^2 \beta+(1+r^2)f\}\gamma^2
]\ ,
\label{EOMb}
\end{split}
\\
\begin{split}
\gamma''=&-\frac{1}{2}\bigg[
\frac{(fg\beta)'}{fg\beta}
-\frac{2 \alpha'}{\alpha} 
+\frac{2(1+3r^2)}{r(1+r^2)}\bigg]\gamma'
-\frac{4\alpha^2 (r^2h^2 \beta -(1+r^2)f)}{r^2 (r^2+1)^2 f g \beta}\gamma
\ .
\end{split}
\label{EOMc}
\end{align}
We will solve the coupled nonlinear ordinary differential equations numerically.
Because of the $SU(2)$-symmetric ansatz, the problem of solving the Einstein-Maxwell equations is reduced to the one-dimensional problem.

\subsection{Physical quantities at the AdS boundary}

From the asymptotic behavior of the fields near the AdS boundary, we can obtain the boundary stress tensor $T_{ij}$ and current $j_i$.
For the boundary condition, we assume that the spacetime is asymptotically AdS and there is no external electromagnetic field at the AdS boundary: 
\begin{equation}
 f,\alpha,\beta\to 1\ ,\quad 
 \gamma\to 0\quad (r\to \infty)\ .
\label{asymAdS}
\end{equation}
Note that $g\to 1$ follows from the above conditions and the equations of motion~(\ref{EOMf}-\ref{EOMc}).
Solving the equations of motion near the AdS boundary, 
we obtain the asymptotic form of the fields as
\begin{equation}
\begin{split}
&f(r)=1+\frac{c_f}{r^4}+\cdots\ ,\quad
g(r)=1+\frac{c_f+c_\beta}{r^4}+\cdots\ ,\quad
h(r)=\Omega+\frac{c_h}{r^4}+\cdots\ ,\\
&\alpha(r)=1+\frac{c_\alpha}{r^4}+\cdots\ ,\quad
\beta(r)=1+\frac{c_\beta}{r^4}+\cdots\ ,\quad
\gamma=\frac{j}{2r^2}+\cdots\ ,
\end{split}
\label{asym}
\end{equation}
where $\Omega$, $c_k$ ($k=f,g,h,\alpha,\beta$), and $j$ are constants. 
Note that, as we assume $\alpha \neq 1$, we can no longer freely shift the value of $\Omega$ while keeping the form of the metric~(\ref{metricanz}).
The asymptotic form of the metric is given by
\begin{equation}
 ds^2\simeq -(1+r^2)d\tau^2 + \frac{dr^2}{1+r^2}+\frac{r^2}{4} \left[ \sigma_1^2+\sigma_2^2+(\sigma_3+2\Omega d\tau)^2 \right] \ .
\end{equation}

The ansatz~(\ref{metricanz}) actually corresponds to the rotating frame at infinity.
We can move to the non-rotating frame by the coordinate transformation~(\ref{psishift}).
In this frame, the Killing vector $\partial_\tau$ is written as
\begin{equation}
K = \partial_\tau = \partial_t + \Omega \partial_{\psi/2}\ .
\label{killingK}
\end{equation}
This can be regarded a helical Killing vector with respect to $(t,\psi)$.
In the rotating frame at infinity, one might interpret that the rotational $U(1)$ symmetry for $\chi$ is simply broken, while in the non-rotating frame the interpretation is that the time translations and rotations generated by $\partial_t$ and $\partial_\psi$ are broken to the helical subgroup as (\ref{killingK}).

The constants $\Omega$ and $c_k$  
are related to the boundary stress tensor $T_{ij}$ ($i,j=t,\theta,\phi,\psi$) as
\begin{align}
 8\pi G_5 T_{ij}dx^i dx^j
&= \frac{1}{2}(c_\beta-3c_f) d\tau^2
+2 c_h d\tau(\sigma_3+2\Omega d\tau)
-\frac{c_f+c_\beta}{8}(\sigma_1^2+\sigma_2^2)\nonumber \\
&\hspace{2cm}+\frac{c_\alpha}{2}(\sigma_1^2-\sigma_2^2)
+\frac{1}{8}(-c_f+3c_\beta)(\sigma_3+2\Omega d\tau)^2\ .
\end{align}
See Ref.~\cite{Ishii:2018oms} for the detail of the derivation.
The boundary electric current $j_i$ is defined by
\begin{equation}
j_i \equiv \frac{\gamma_{ij}}{\sqrt{-\gamma}}\frac{\delta S}{\delta A^\textrm{can}_j}\ ,
\end{equation}
where $A^\textrm{can}_j$ is the boundary value of the canonically normalized Maxwell field, 
$A^\textrm{can}_i=A_i/\sqrt{16\pi G_5}|_{r=\infty}$,
and $\gamma_{ij}$ is the boundary metric: $\gamma_{ij} dx^i dx^j = -d\tau^2 + \{\sigma_1^2+\sigma_2^2+(\sigma_3+2\Omega d\tau)^2\}/4$.
The constant $j$ corresponds to the electric current as
\begin{equation}
 \sqrt{16\pi G_5}\, j_i dx^i = j \sigma_1\ .
\label{currentdef}
\end{equation}
We will ignore the irrelevant prefactor $\sqrt{16\pi G_5}$ and simply call $j$ as the electric current.

To discuss physical conserved charges, it is convenient to work in the non-rotating frame at infinity.
In that frame, the asymptotic form of the metric becomes 
\begin{equation}
 ds^2\simeq -(1+r^2)dt^2 +\frac{dr^2}{1+r^2} +\frac{r^2}{4}(\bar{\sigma}_1^2+\bar{\sigma}_2^2+\bar{\sigma}_3^2)\ .
\end{equation}
The boundary stress tensor and electric current are rewritten as
\begin{align}
 8\pi G_5 T_{ij}dx^i dx^j
&= \frac{1}{2}(c_\beta-3c_f) dt^2
+2 c_h dt \bar{\sigma}_3-\frac{c_f+c_\beta}{8}(\bar{\sigma}_1^2+\bar{\sigma}_2^2)\nonumber \\
&\qquad +c_\alpha (e^{4i\Omega t}\bar{\sigma}_+^2+e^{-4i\Omega t}\bar{\sigma}_-^2)
+\frac{1}{8}(-c_f+3c_\beta)\bar{\sigma}_3^2
\ ,\\
\sqrt{16\pi G_5} j_i dx^i &= j(e^{2i\Omega t} \bar{\sigma}_+ + e^{- 2i\Omega t} \bar{\sigma}_-)\ .
\label{Tmunu_non-rot}
\end{align}
The energy density ($\propto T_{tt}$) and angular momentum density ($\propto T_{t\psi}$) 
depend on neither time nor spatial coordinates.
Therefore, the energy and angular momentum are constant in time and given by
\begin{equation}
E=\int d\Omega_3 T_{tt} =\frac{\pi(c_\beta-3c_f)}{8 G_5}\ ,\quad
J=-\int d\Omega_3 T_{t (\psi/2)} =-\frac{\pi c_h}{2 G_5}\ .
\label{EJdef}
\end{equation}
On the other hand, the stress part of the boundary stress tensor (the coefficients of $\bar{\sigma}_{\pm}$) and
the electric current depend on the asymptotic time $t$.
The dependence is time periodic.
Hence, the photonic black resonators are considered as time periodic solutions in terms of the time $t$ in the non-rotating frame.
From the viewpoint of the AdS/CFT correspondence, 
black resonators would be considered to be dual to time periodic states in the boundary field theory.

For notational simplicity, we will set $G_5=1$ when we show our numerical results.
We can easily recover $G_5$ by replacing $E\to G_5 E$,  $J\to G_5 J$, etc.

\section{Photon stars}
\label{phstr}

In this section, we consider horizonless ``geon'' solutions in the presence of a nontrivial Maxwell field.
We refer to them as {\it photon stars}.
In the absence of the black hole horizon, the boundary condition is imposed at the center of the global AdS $r=0$ and different from that for black resonators.
Therefore, we separately discuss the horizonless solutions in this section.

\subsection{Perturbative construction}

We start by constructing photon stars perturbatively near pure AdS background.
We consider the perturbative expansion of the solution as
\begin{equation}
\Phi(r) = \sum_{m=0}^{\infty} \Phi^{(m)}(r) \epsilon^m \ ,
\label{ps_pert}
\end{equation}
where $\Phi(r) = (f(r),g(r),h(r),\alpha(r),\beta(r),\gamma(r))^T$ collectively denotes the fields, and $\epsilon$ is a small parameter.
We substitute (\ref{ps_pert}) into (\ref{EOMf}-\ref{EOMc}) and expand them in powers of $\epsilon$.
Resulting equations are solved perturbatively.

The lowest order $m=0$ is the background solution given by the global AdS$_5$ with the trivial gauge field in the rotating frame: $\Phi^{(0)}(r)=(1,1,\Omega^{(0)},1,1,0)^T$.
We set $\Omega^{(0)}=2$ for the onset of the fundamental tone $n=0$.
In the leading order $m=1$, we turn on $\gamma(r)$ as the contribution triggering a photon star:\footnote{Alternatively, a purely gravitational perturbation by $\alpha^{(1)} \neq 0$ while $\gamma^{(1)} =0$ could be also turned on at $\Omega^{(0)}=2$, but this does not result in a solution with a nontrivial $\gamma^{(m)}$. We also find that we cannot simultaneously turn on both $\alpha^{(1)}$ and $\gamma^{(1)}$ in the leading order for a trigger because doing that results in inconsistency in higher orders. Therefore, one of them must be zero.} $\Phi^{(1)}(r)=(0,0,0,0,0,\gamma^{(1)}(r))^T$. The solution at $m=1$ is nothing but (\ref{linear_gamma_sol}), which at $\Omega=2$ becomes
\begin{equation}
\gamma^{(1)}(r) = \frac{r^2}{(1+r^2)^2} \ .
\end{equation}

The nonlinear perturbation to higher orders can be continued in a similar way as in \cite{Ishii:2018oms}.
To solve the perturbative equations of motion, we impose regularity at $r=0$ and asymptotically AdS behavior in $r \to \infty$ with a trivial gauge field $\gamma^{(m)}(r)|_{r=\infty} = 0$.
We obtain the solution as
\begin{equation}
\begin{split}
f(r) &= 1+ f^{(2)}(r) \epsilon^2 + f^{(4)}(r) \epsilon^4 + \mathcal{O}(\epsilon^6)\ , \\
g(r) &= 1+ g^{(2)}(r) \epsilon^2 + g^{(4)}(r) \epsilon^4 + \mathcal{O}(\epsilon^6)\ , \\
h(r) &= 2 + h^{(2)}(r) \epsilon^2 + h^{(4)}(r) \epsilon^4 + \mathcal{O}(\epsilon^6)\ , \\
\alpha(r) &= 1 + a^{(2)}(r) \epsilon^2 + a^{(4)}(r) \epsilon^4 + \mathcal{O}(\epsilon^6)\ , \\
\beta(r) &= 1 + \beta^{(2)}(r) \epsilon^2 + \beta^{(4)}(r) \epsilon^4 + \mathcal{O}(\epsilon^6)\ , \\
\gamma(r) &= \gamma^{(1)}(r) \epsilon + \gamma^{(3)}(r) \epsilon^3 + \gamma^{(5)}(r) \epsilon^5 + \mathcal{O}(\epsilon^6) \ ,
\end{split}
\end{equation}
where
\begin{equation}
\begin{split}
f^{(2)}(r) &= -\frac{24+84r^2+74r^4+23r^6}{27(1+r^2)^5}\ , \quad
g^{(2)}(r) = -\frac{r^2(66+5r^2+20r^4)}{27(1+r^2)^5}\ , \\
h^{(2)}(r) &= -\frac{1031+573r^2+258r^4+86r^6}{945(1+r^2)^3}\ , \\
\alpha^{(2)}(r) &= -\frac{2r^2(-223+1380r^2)}{4851(1+r^2)^4}\ , \quad
\beta^{(2)}(r) = \frac{r^2(10+r^2)}{9(1+r^2)^4}\ , \\
\gamma^{(3)}(r) &= -\frac{r^4(-703672+343582r^2+1055778r^4+778037r^6+187393r^8)}{873180(1+r^2)^7}\ , \quad
\end{split}
\end{equation}
while we do not reproduce the lengthy expressions in $m \ge 4$.
It is important that a nontrivial $\gamma^{(1)}$ induces a nonzero $\alpha^{(2)}$.
In this way, the nontrivial gauge field deforms the global AdS geometry into a time periodic photon star.
Up to $\mathcal{O}(\epsilon^5)$, the physical quantities are obtained as
\begin{equation}
\begin{split}
E &= \pi \left( \frac{1}{3} \epsilon^2 - \frac{403335607}{7059660300} \epsilon^4 \right)\ , \quad
J = \pi \left( \frac{1}{6} \epsilon^2 - \frac{349796737}{14119320600} \epsilon^4 \right)\ , \\
\Omega &= 2 - \frac{86}{945} \epsilon^2 + \frac{102955247}{1111896497250} \epsilon^4 \ , \quad
j = 2 \epsilon - \frac{187393}{436590} \epsilon^3 \ .
\end{split}
\label{perturb_EJOmegaj}
\end{equation}

\subsection{Full nonlinear solution}
To construct photon stars away from the pure AdS, we numerically solve the equations of motion (\ref{EOMf}-\ref{EOMc}).
Imposing boundary conditions at $r=0$ and $r \to \infty$, we use the shooting method.
On the one hand, evaluating near the center of the AdS $(r=0)$ and requiring regularity, we obtain series solutions,
\begin{equation}
\begin{split}
 f(r)&=f_0 + \frac{4}{3}f_0 \gamma_2^2 r^2 +\mathcal{O}(r^4)\ ,\quad 
 g(r)=1 -(\beta_2+\frac{4}{3}\gamma_2^2)r^2+\mathcal{O}(r^4)\ ,\\
h(r)&=h_0+\frac{4}{3}h_0 \gamma_2^2 r^2 +\mathcal{O}(r^4)\ ,\quad
\alpha(r)=1+\alpha_2 r^2 +\mathcal{O}(r^4)\ ,\\
\beta(r)&=1+\beta_2 r^2 +\mathcal{O}(r^4)\ ,\quad
\gamma(r)=\gamma_2 r^2 +\mathcal{O}(r^4)\ .
\end{split}
\end{equation}
There are five free parameters in the above expression: $(f_0, h_0, \alpha_2, \beta_2, \gamma_2)$.
On the other hand, at $r=\infty$, we need to satisfy the four conditions in~(\ref{asymAdS}).
Therefore, the photon stars are in a one-parameter family.
To obtain a solution, in our numerical calculations, 
we fix the value of $\gamma_2$ and tune the other four parameters $(f_0, h_0, \alpha_2, \beta_2)$ 
by the shooting method so that (\ref{asymAdS}) is satisfied. 
We repeat this procedure by varying $\gamma_2$ and construct the family of photon stars.

\begin{figure}[t]
  \centering
  \subfigure
 {\includegraphics[scale=0.24]{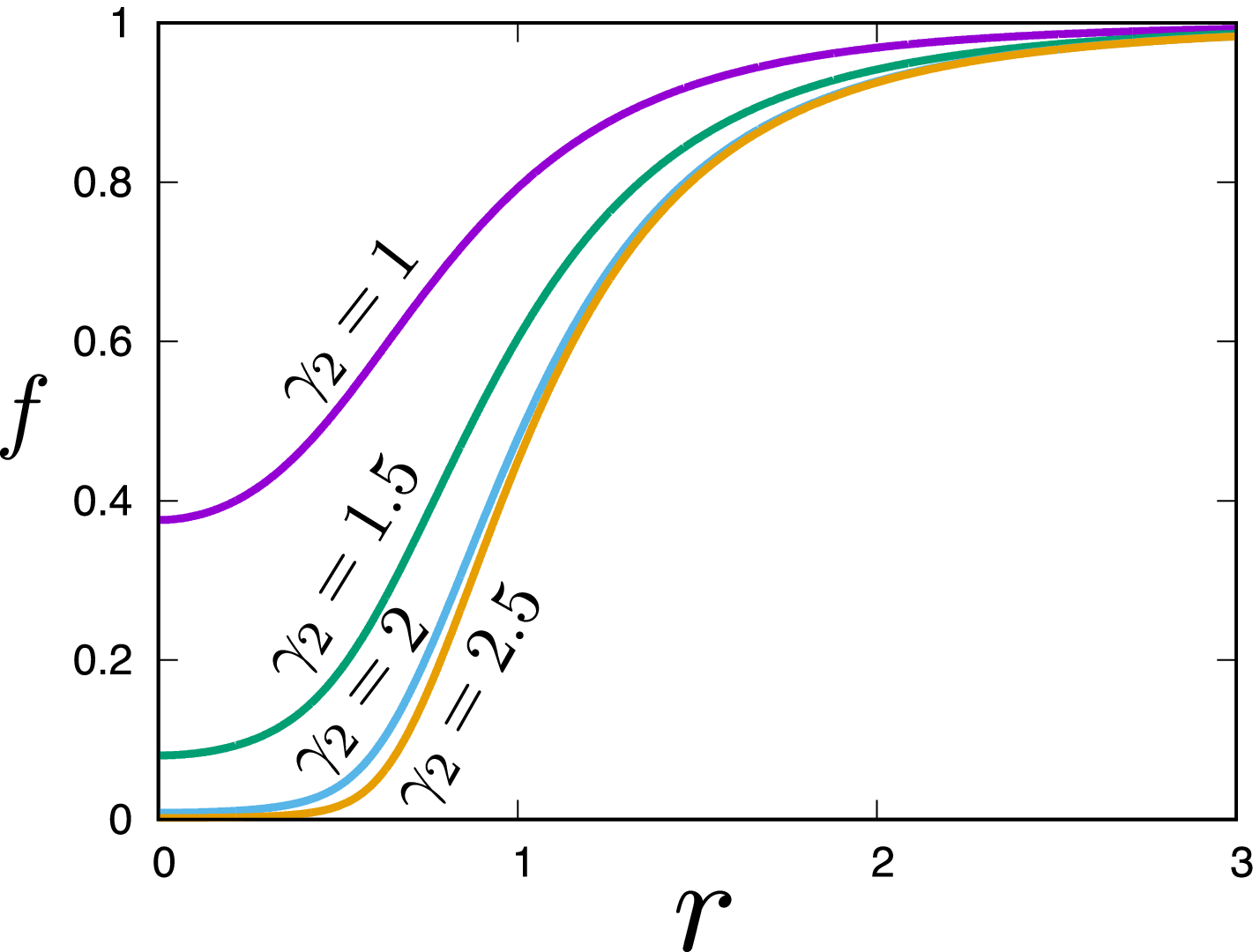}
  }
  \subfigure
 {\includegraphics[scale=0.24]{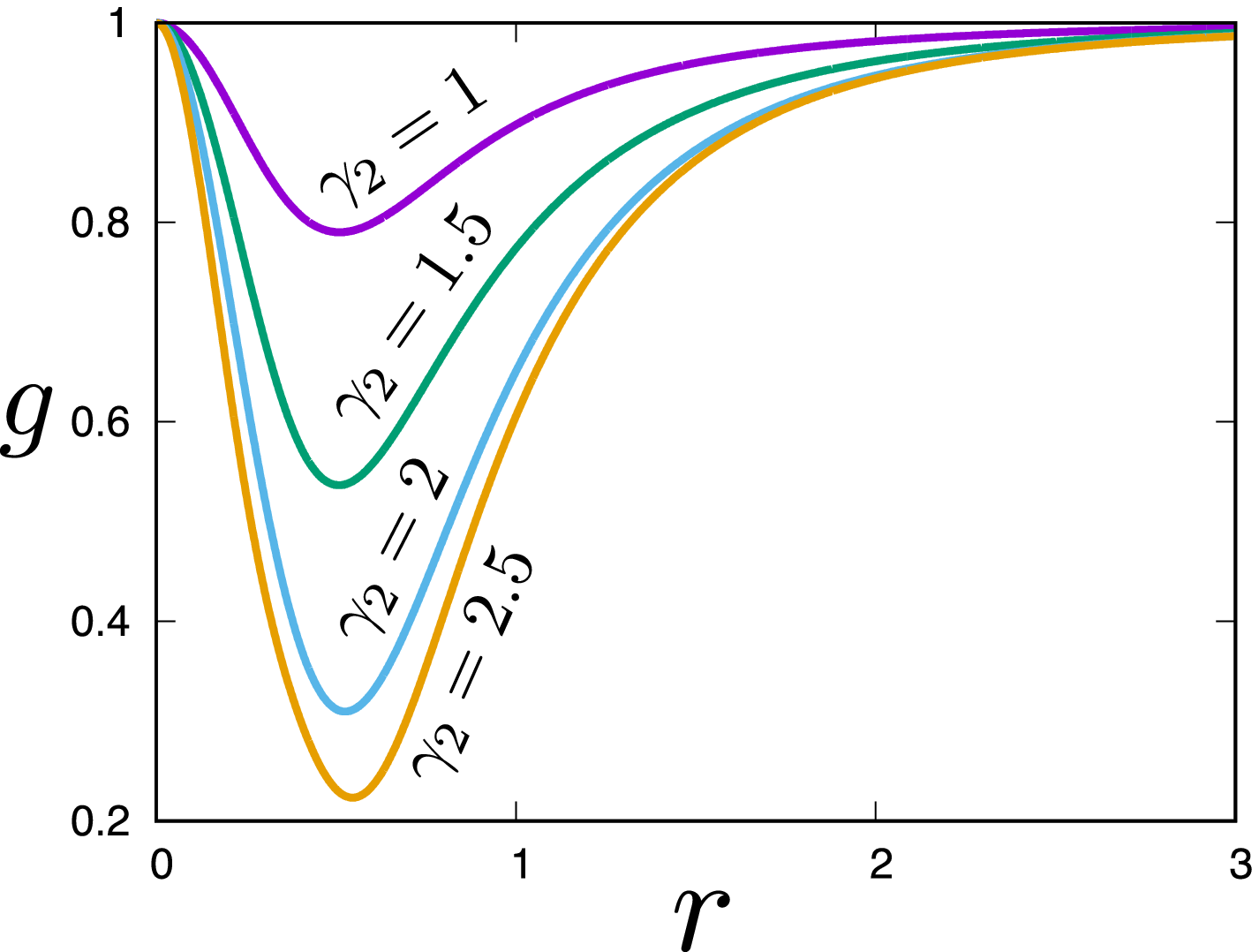}
  }
  \subfigure
 {\includegraphics[scale=0.24]{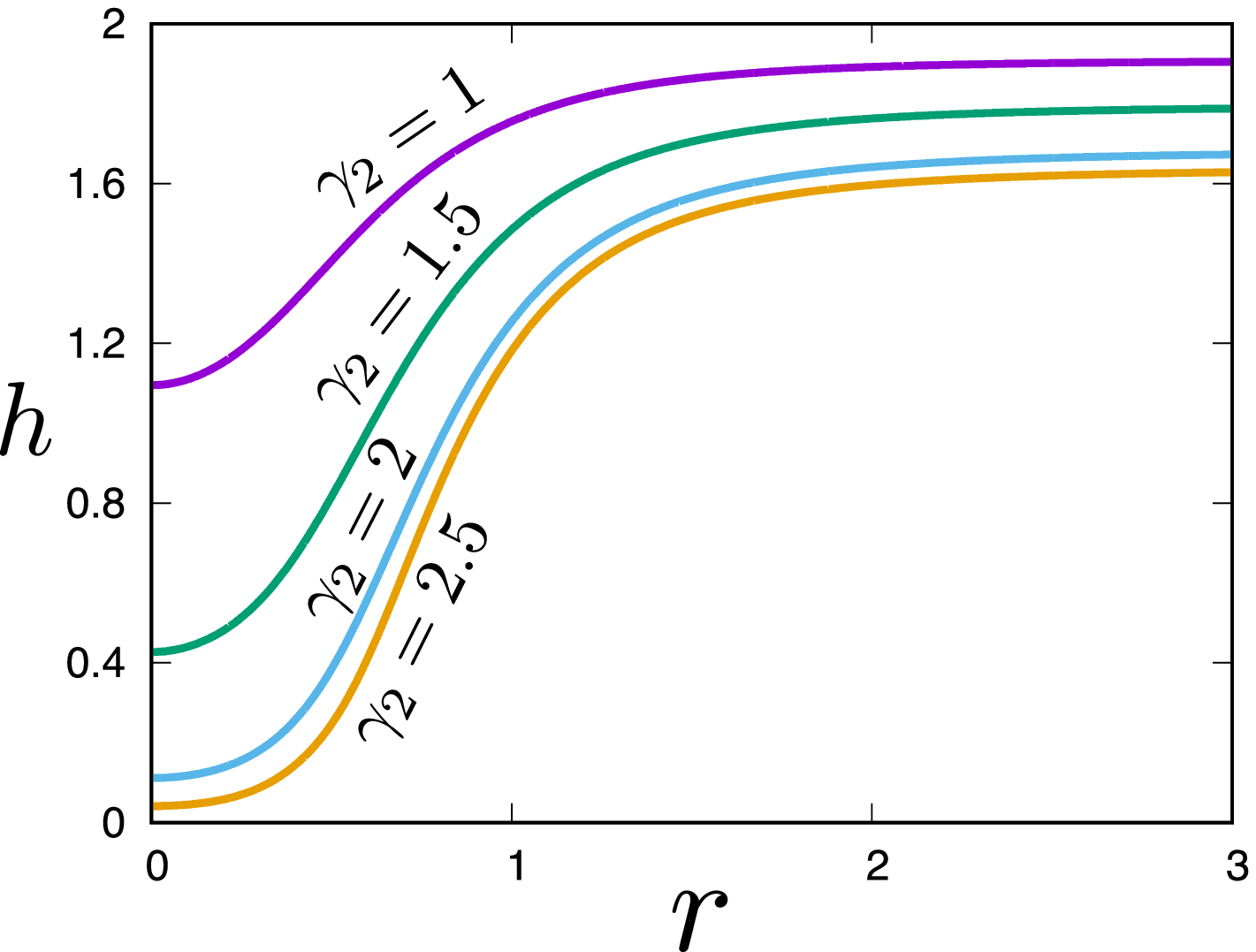}
  }
  \subfigure
 {\includegraphics[scale=0.24]{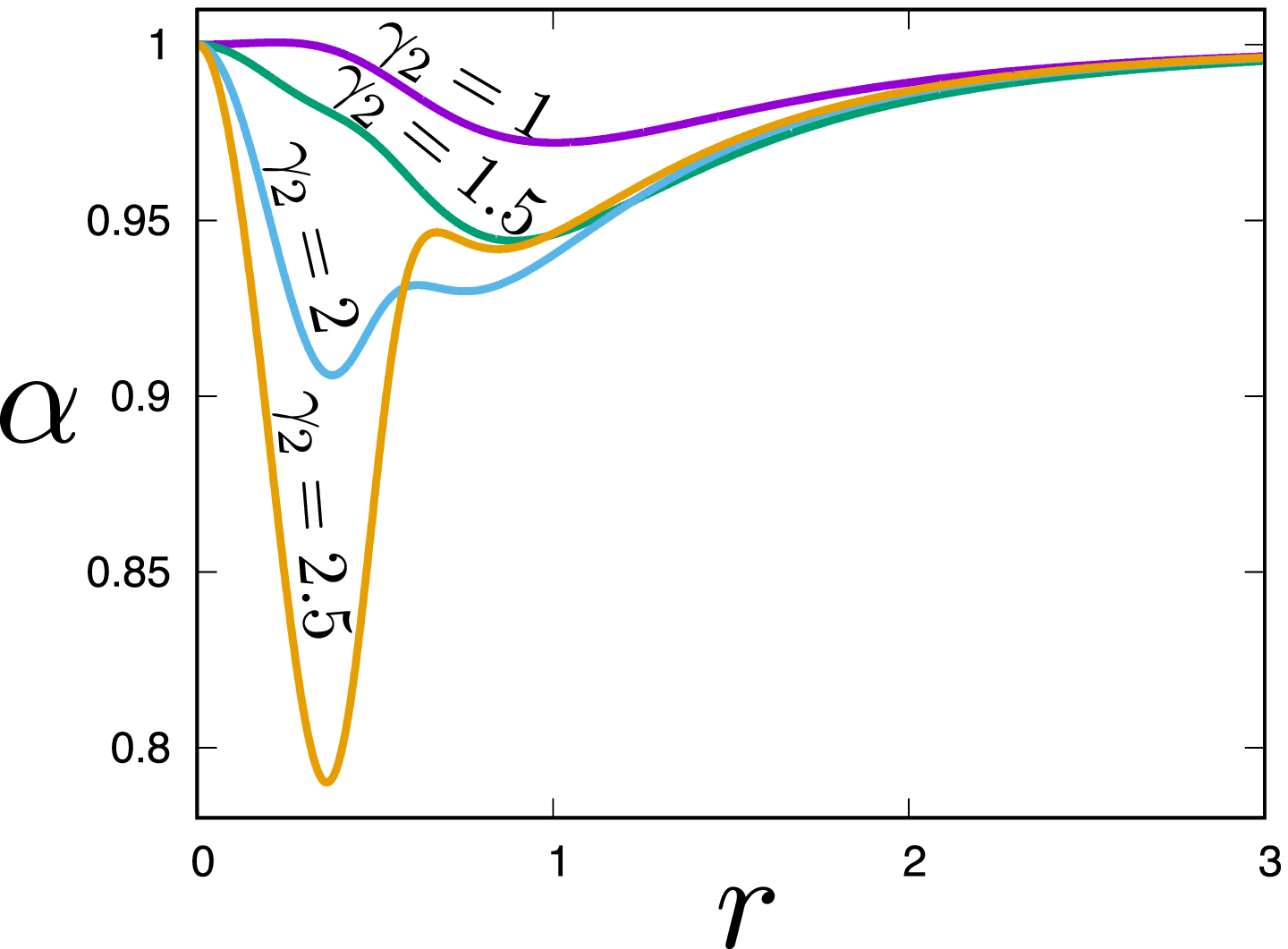}
  }
  \subfigure
 {\includegraphics[scale=0.24]{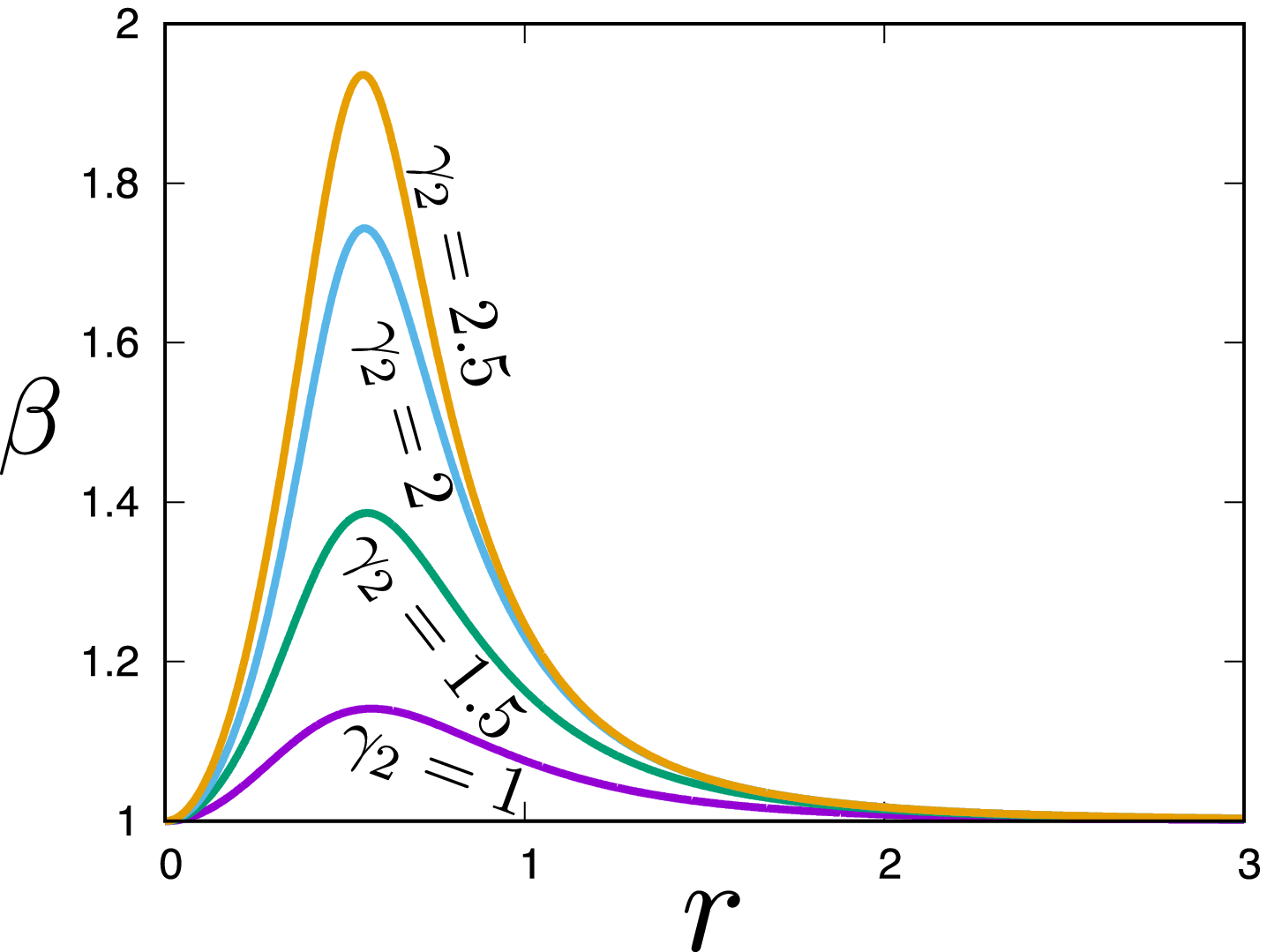}
  }
  \subfigure
 {\includegraphics[scale=0.24]{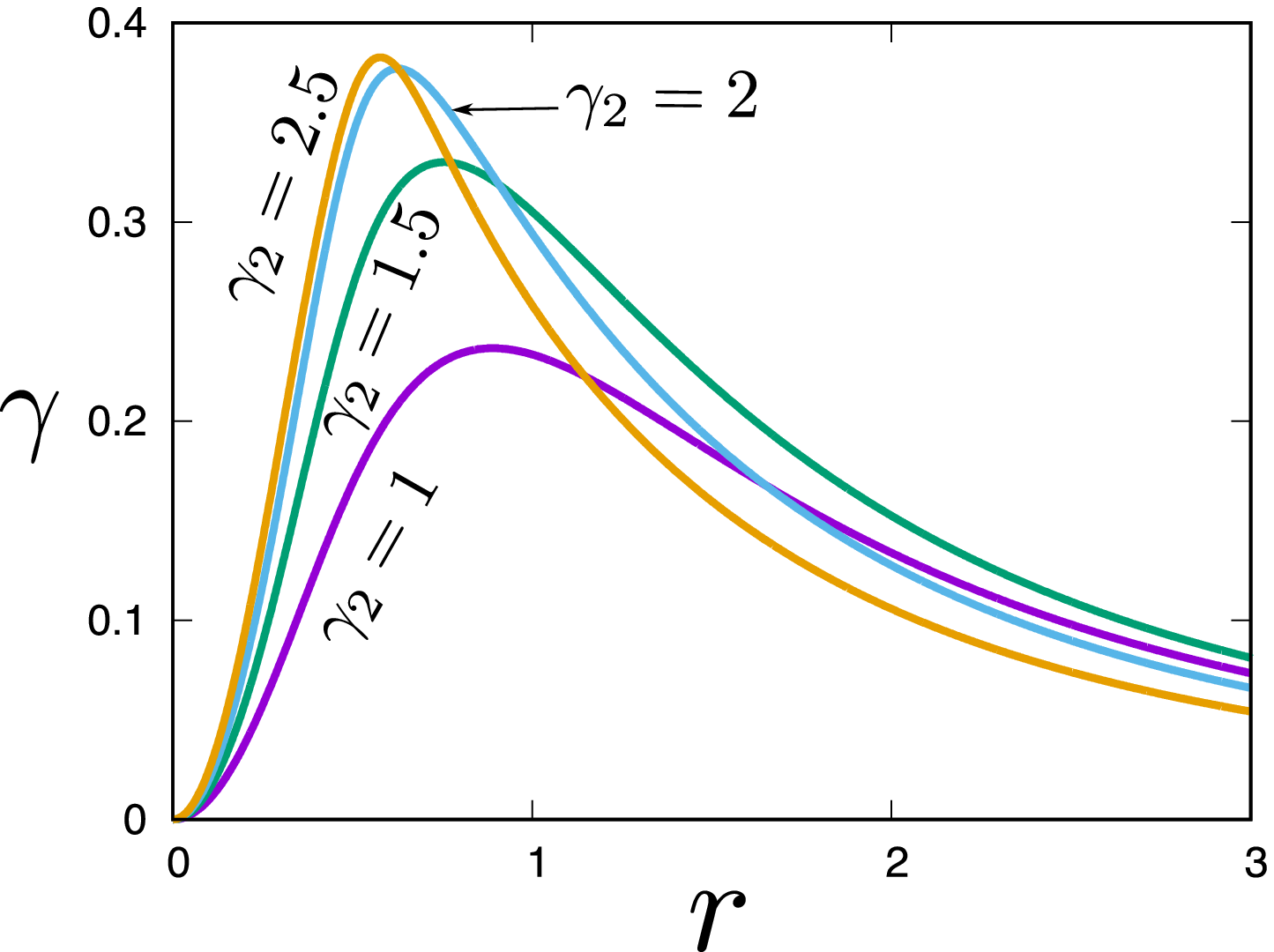}
  }
\subfigure
 {\includegraphics[scale=0.24]{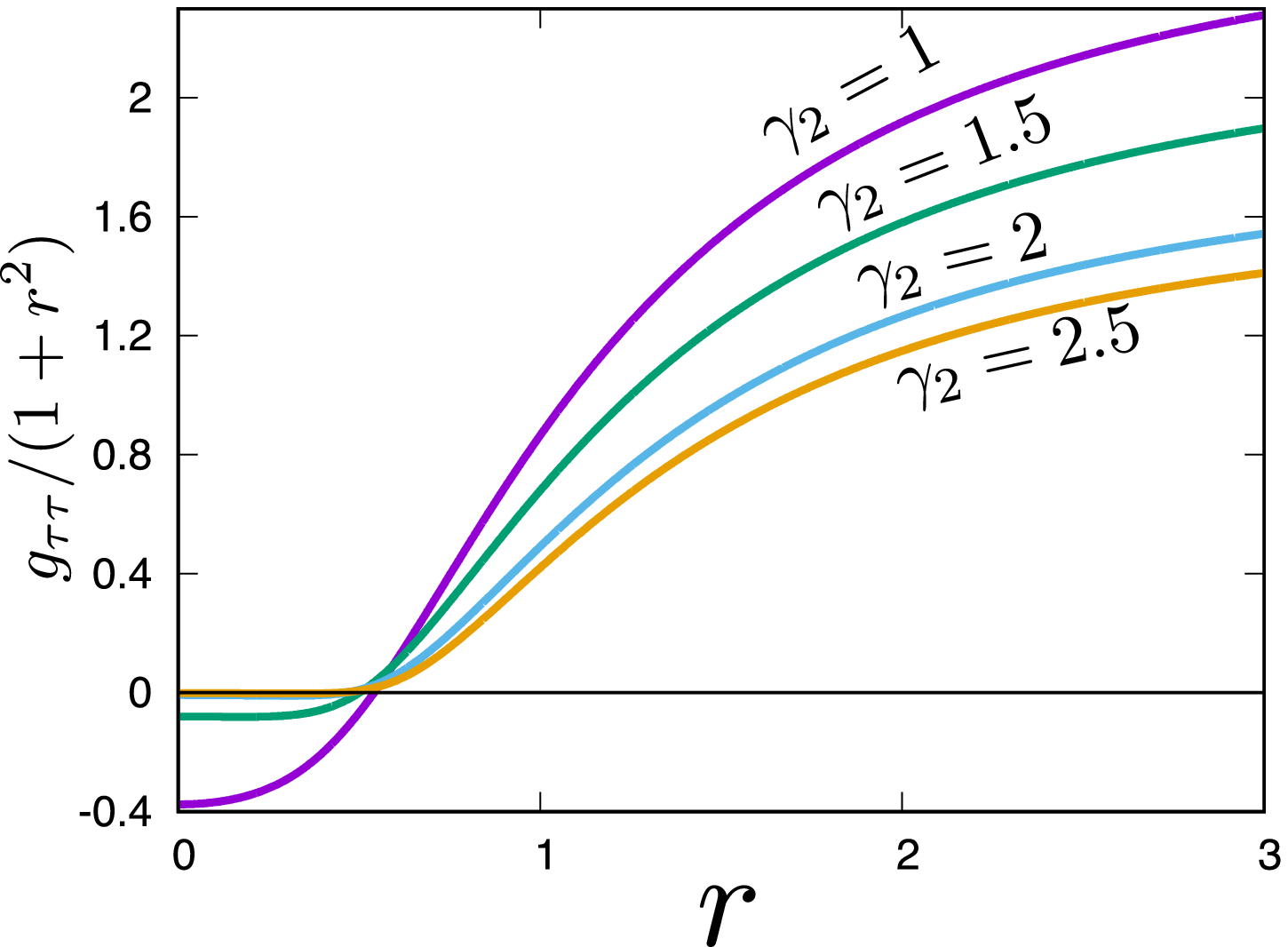}
  }
 \caption{%
Profile of $f,g,h,\alpha,\beta$ and $\gamma$ for photon stars with $\gamma_2=1,1.5,2,2.5$.
The norm of the helical Killing vector, $K^2=g_{\tau\tau}$, is also shown, 
which is normalized by $1+r^2$ for visibility. 
}
\label{photonstarfncs}
\end{figure}

In Fig.~\ref{photonstarfncs}, we show the profile of the functions $f,g,h,\alpha,\beta$ and $\gamma$ for $\gamma_2=1,1.5,2,2.5$. 
Note that $\gamma_2=0$ corresponds to the pure AdS.
As $\gamma_2$ increases, the spacetime is deformed from the pure AdS.
Shown in the last panel is the norm of the Killing vector (normalized by a function of $r$), $g_{\mu\nu}K^\mu K^\nu/(1+r^2)$.
It is timelike near $r=0$ but becomes spacelike near the infinity.
The latter implies that the spacetime of the photon star is dynamical.

\begin{figure}[t]
  \centering
  \subfigure[Mass]
 {\includegraphics[scale=0.33]{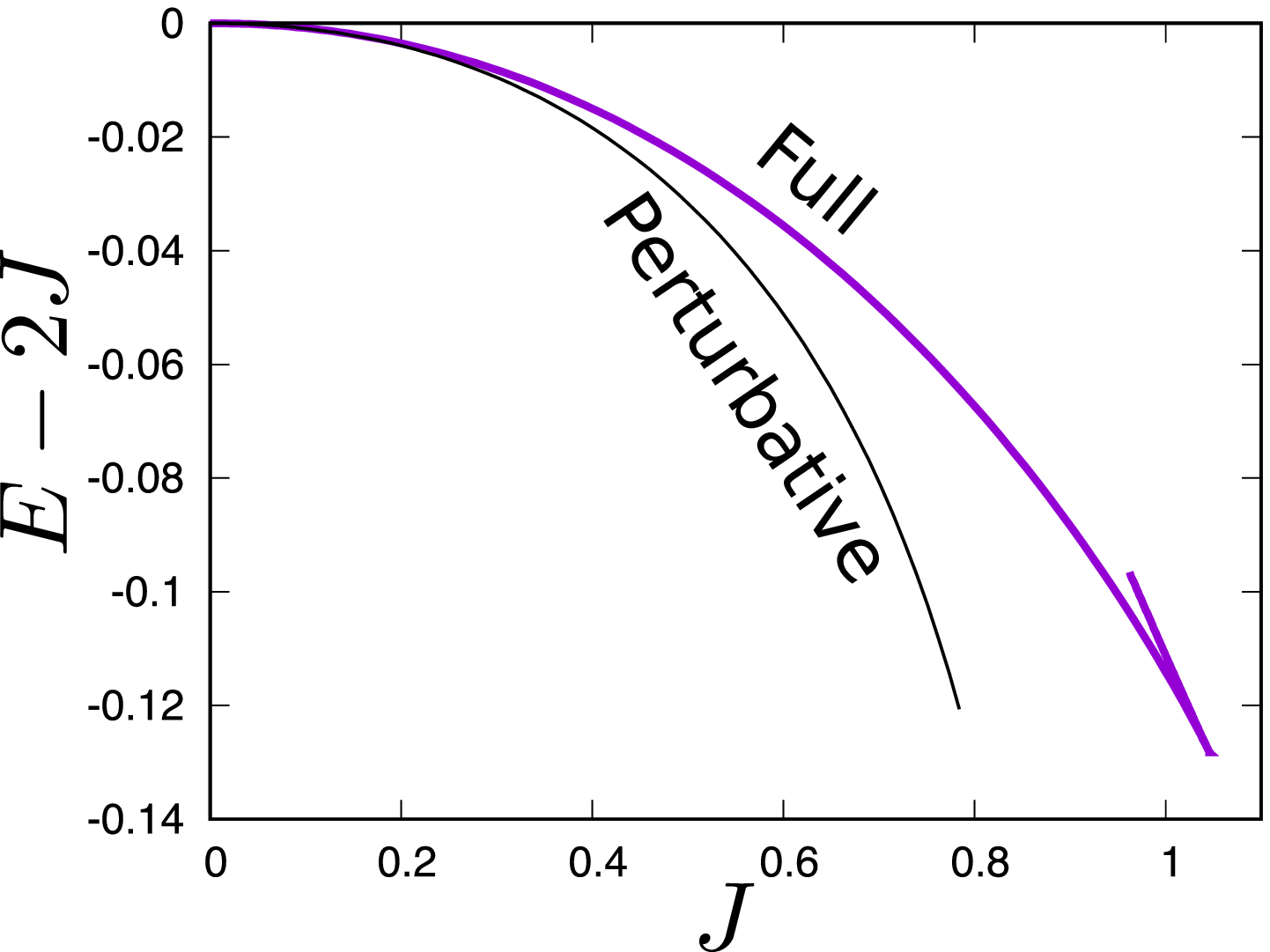}\label{fig_J_E_n0_phstr}
  }
  \subfigure[Angular velocity]
 {\includegraphics[scale=0.33]{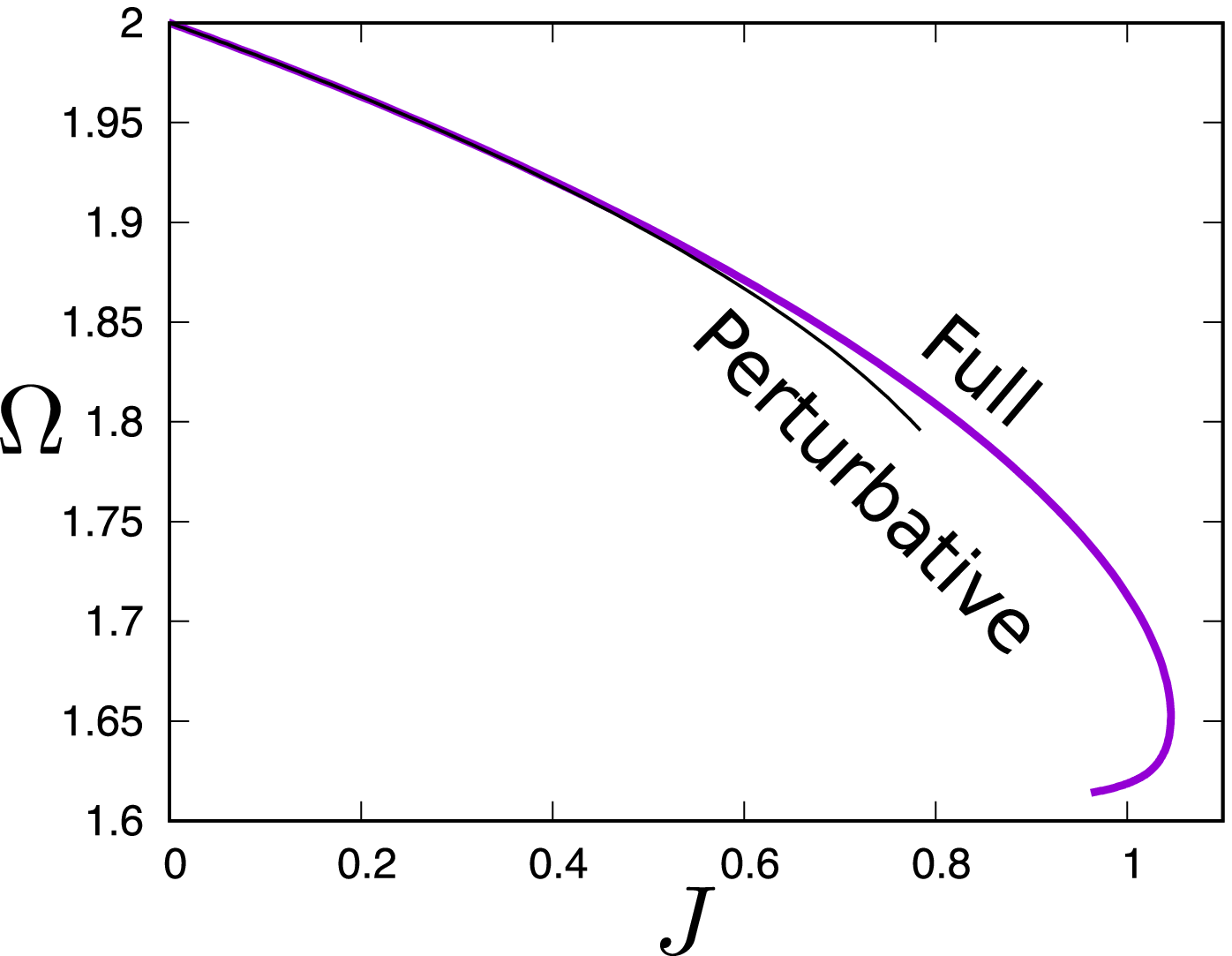}
  }
  \subfigure[Current]
 {\includegraphics[scale=0.33]{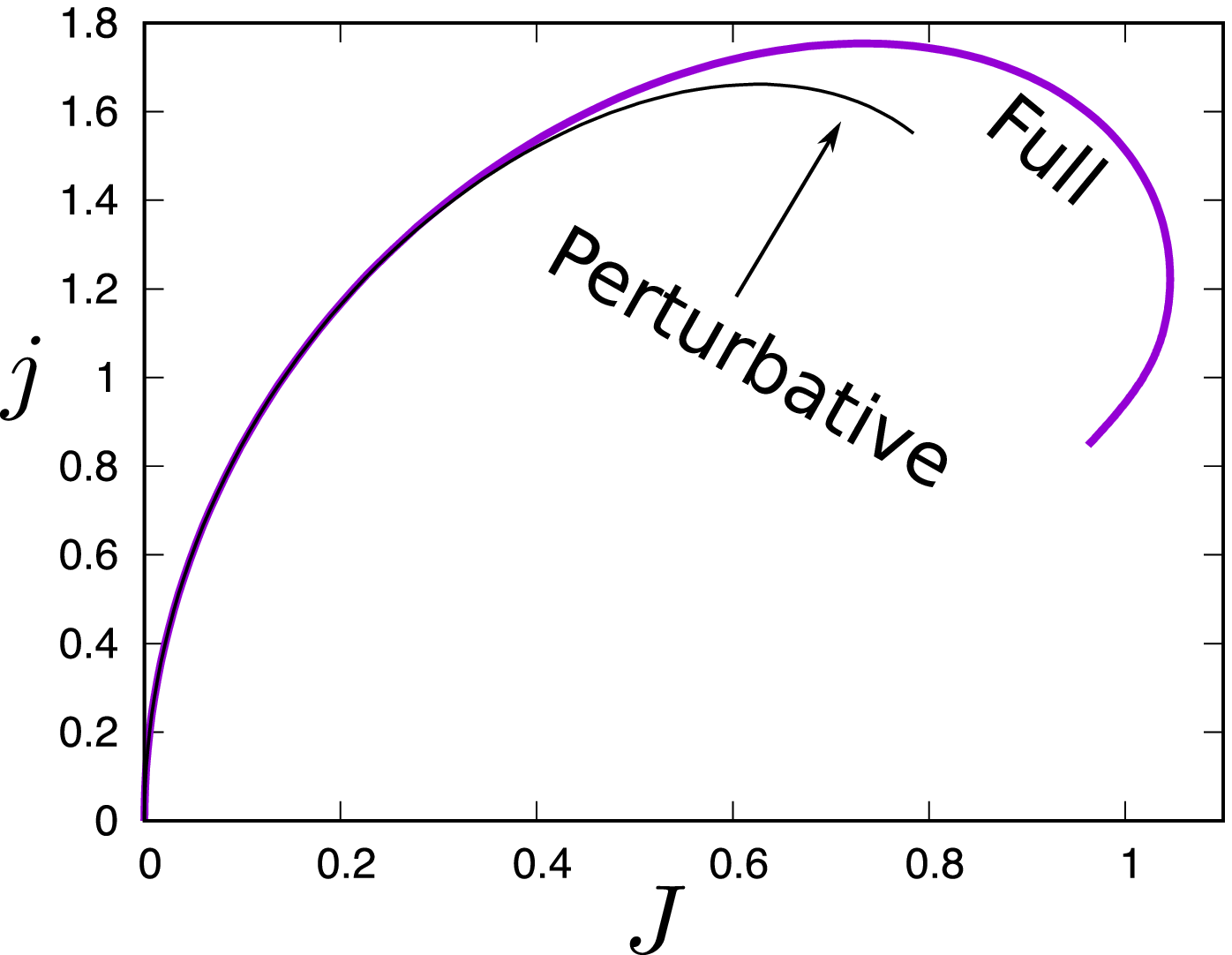}
  }
 \caption{%
Mass $E$, angular velocity $\Omega$, and electric current $j$ of photon stars 
as a function of the angular momentum $J$.
The perturbative results (\ref{perturb_EJOmegaj}) are also plotted for $\epsilon \le 1.5$ for comparison.
For visibility, $E-2J$ is taken in the plot of the mass.
In each diagram, there is a turning point where $J$ is maximum.}
\label{photonstarJE}
\end{figure}

From the asymptotic behavior of the numerical solutions, 
we can read out the coefficients $\Omega$, $c_k$ and $j$ in (\ref{asym}).
They are converted into the physical quantities of photon stars by (\ref{currentdef}) and (\ref{EJdef}).
In Fig.~\ref{photonstarJE}, we plot the mass $E$, angular velocity $\Omega$, and electric current $j$ 
of the photon star as a function of the angular momentum $J$.
For visibility, we take $E-2J$ as the vertical axis in the plot of the mass.
The results for full numerical and perturbative solutions are shown in purple and black curves, respectively.
We can see a good agreement of them in $J\lesssim 0.5$.
Here, we only focus on the photon stars for the fundamental tone $n=0$ bifurcating from $\Omega=2$ in the pure AdS. 
Overtones are studied in appendix \ref{overtone}.

It is remarkable that each diagram has a ``turning point'' where the value of $J$ reaches its upper limit.
The physical quantities of photon stars are multivalued around the maximum $J$.
In particular, the mass should form a cusp at the maximum $J$ because of the first law: $dE=\Omega dJ$.
This is actually seen in Fig.~\ref{fig_J_E_n0_phstr}.
In comparison, in the geons studied in Ref.~\cite{Ishii:2018oms}, we did not find such turning points as long as we constructed solutions within numerical limitations.

In the photon star solutions we constructed, the angular velocity always satisfies $\Omega>1$ although it decreases from the value $\Omega=2$ of the pure AdS.
This again implies that the spacetime of the photon star is dynamical.

\section{Photonic black resonators}
\label{phtoBR}

In this section, we study the cohomogeneity-1 photonic black resonator solutions given by our ansatz~(\ref{Mxwpert}) and (\ref{metricanz}).

To solve the equations of motion (\ref{EOMf}-\ref{EOMc}) numerically, we need to decide how to fix the constants at the boundaries of the computational domain $r_h \le r < \infty$.
We impose the metric functions $f(r)$ and $g(r)$ to be zero at the horizon $r=r_h$.
Then, solving (\ref{EOMf}-\ref{EOMc}) near the horizon, we obtain the asymptotic solution as
\begin{equation}
\begin{split}
f(r)&=f_1(r-r_h)+\cdots\ ,\quad
g(r)=g_1(r-r_h)+\cdots\ ,\\
h(r)&=h_0 + h_1(r-r_h)+\cdots\ ,\quad
\alpha(r)=\alpha_0 + \alpha_1(r-r_h)+\cdots\ ,\\
\beta(r)&=\beta_0 + \beta_1(r-r_h)+\cdots\ ,\quad
\gamma(r)=\gamma_0 + \gamma_1(r-r_h)+\cdots\ ,\quad
\end{split}
\label{phtoBR_Hser}
\end{equation}
where 
\begin{equation}
\begin{split}
&g_1=
\frac{
2\{6 r_h^4 \alpha_0 \beta_0
+3 r_h^2 \beta_0 (\alpha_0^2-\alpha_0 \beta_0 +1)
-2 \alpha_0^2 \gamma_0^2\}}
{3r_h^3  (1+r_h^2)\alpha_0 \beta_0}\ ,
\quad
h_0=0\ ,\\
&\alpha_1
=\frac{
6 \{r_h^2 (\alpha_0^2-1)(\alpha_0^2-\alpha_0 \beta_0 +1)+2 \alpha_0^3 \gamma_0^2\}}
{r_h \{
6 r_h^4 \alpha_0 \beta_0
+3 r_h^2 \beta_0 (\alpha_0^2-\alpha_0 \beta_0 +1)
-2 \alpha_0^2 \gamma_0^2\}}\ ,\\
&\beta_1=-\frac{r_h^2\beta_0^2h_1^2}{f_1(1+r_h^2)}
-\frac{6 \beta_0\{
r_h^2 (
\alpha_0^4
+ \alpha_0^3 \beta_0
-2 \alpha_0^2 \beta_0^2
-2 \alpha_0^2
+ \alpha_0 \beta_0
+1
)
+2 \alpha_0^3 \gamma_0^2
\} }
{ r_h \alpha_0\{6 r_h^4 \alpha_0 \beta_0
+3 r_h^2 \beta_0 (\alpha_0^2-\alpha_0 \beta_0 +1)
-2 \alpha_0^2 \gamma_0^2\} }\ ,\\
&\gamma_1=\frac{6r_h \alpha_0^3 \gamma_0}
{6 r_h^4 \alpha_0 \beta_0
+3 r_h^2 \beta_0 (\alpha_0^2-\alpha_0 \beta_0 +1)
-2 \alpha_0^2 \gamma_0^2}\ .
\end{split}
\label{phtoBR_Hcoeff}
\end{equation}
The condition $h_0=0$ indicates that the helical Killing vector~(\ref{killingK}) is the null generator of the horizon.
It turns out that we have six free parameters at the horizon:
$(r_h,f_1,h_1,\alpha_0,\beta_0,\gamma_0)$.
These are compared with the four conditions~(\ref{asymAdS}) at infinity.
Therefore, we are left with two parameters.
Thus, the photonic black resonator solutions are in a two-parameter family.

In our numerical calculations, 
we specify $(r_h, \gamma_0)$ and tune the other parameters $(f_1,h_1,\alpha_0,\beta_0)$ by the shooting method 
so that the asymptotically AdS conditions~(\ref{asymAdS}) are satisfied.
In this way, we obtain a photonic black resonator solution.
We then repeat this procedure by varying $(r_h, \gamma_0)$.
For a small $\gamma_0$, the geometry of the photonic black resonator is close to that of the MPAdS$_5$.
Hence, we use the value of MPAdS$_5$ for $(f_1,h_1,\alpha_0,\beta_0)$ as the initial guess.
Once the shooting successfully converges, 
we use the resulting values of $(f_1,h_1,\alpha_0,\beta_0)$ 
as the initial guess for the next solution where $\gamma_0$ is slightly varied.

\begin{figure}
  \centering
  \subfigure
 {\includegraphics[scale=0.24]{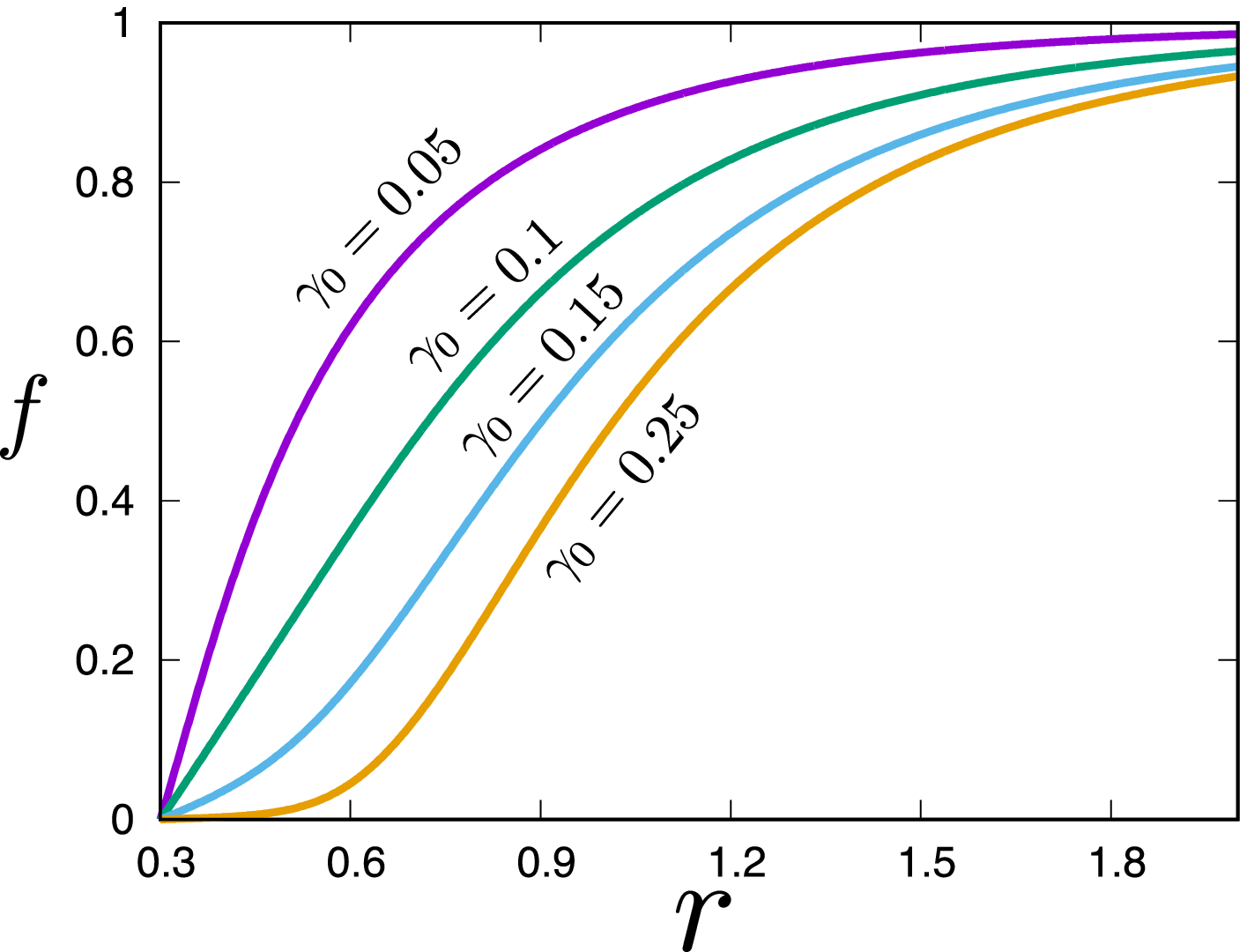}
  }
  \subfigure
 {\includegraphics[scale=0.24]{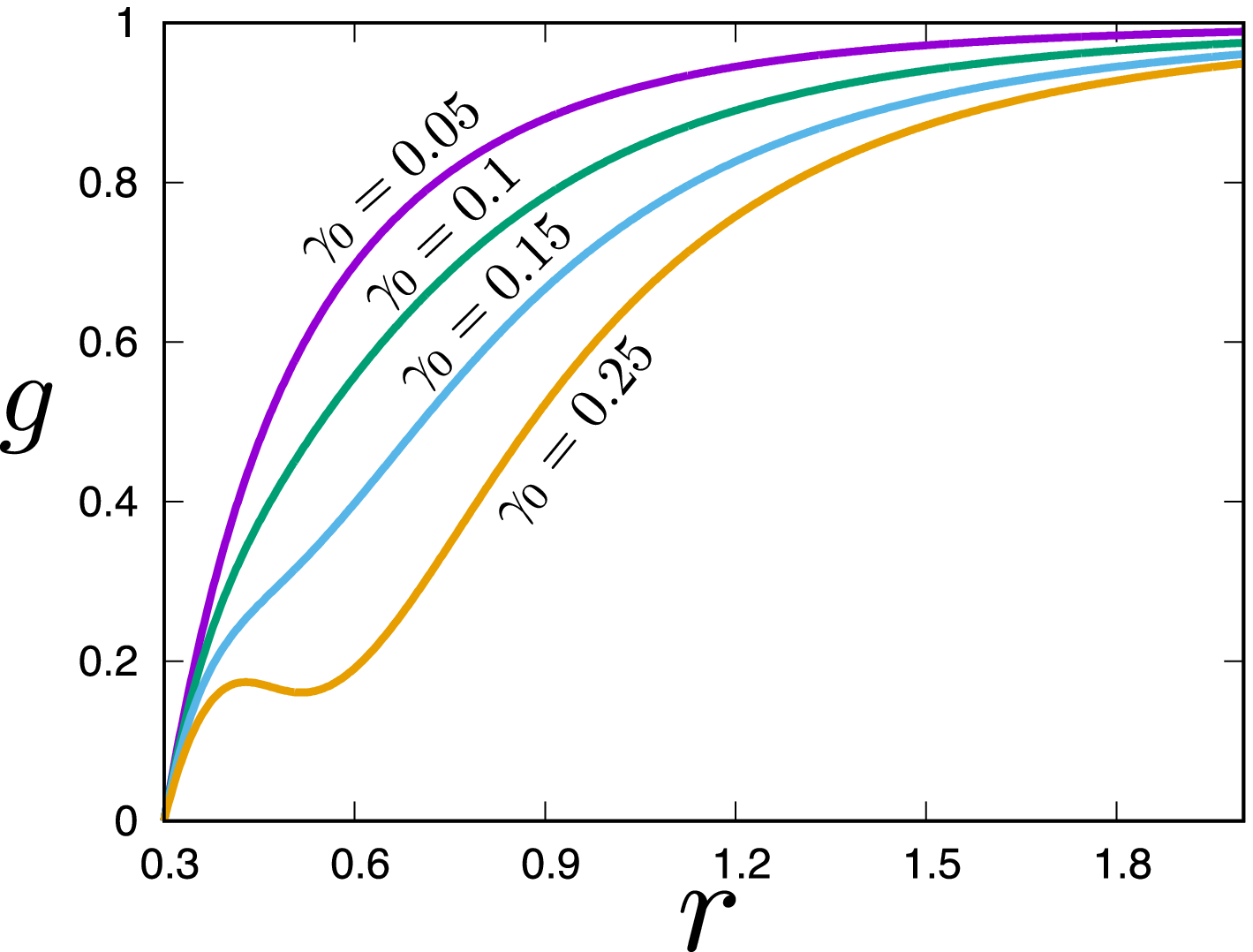}
  }
  \subfigure
 {\includegraphics[scale=0.24]{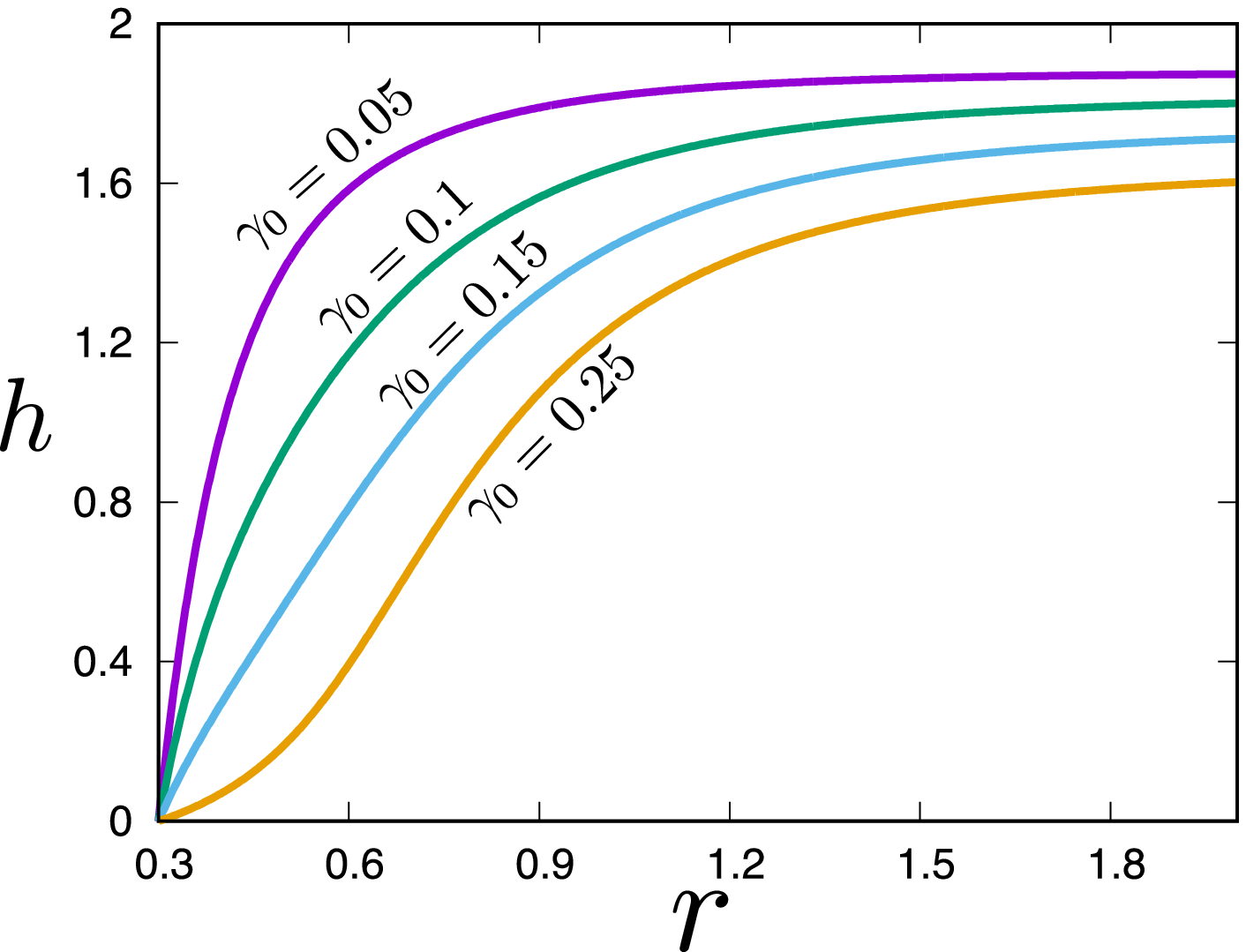}
  }
  \subfigure
 {\includegraphics[scale=0.24]{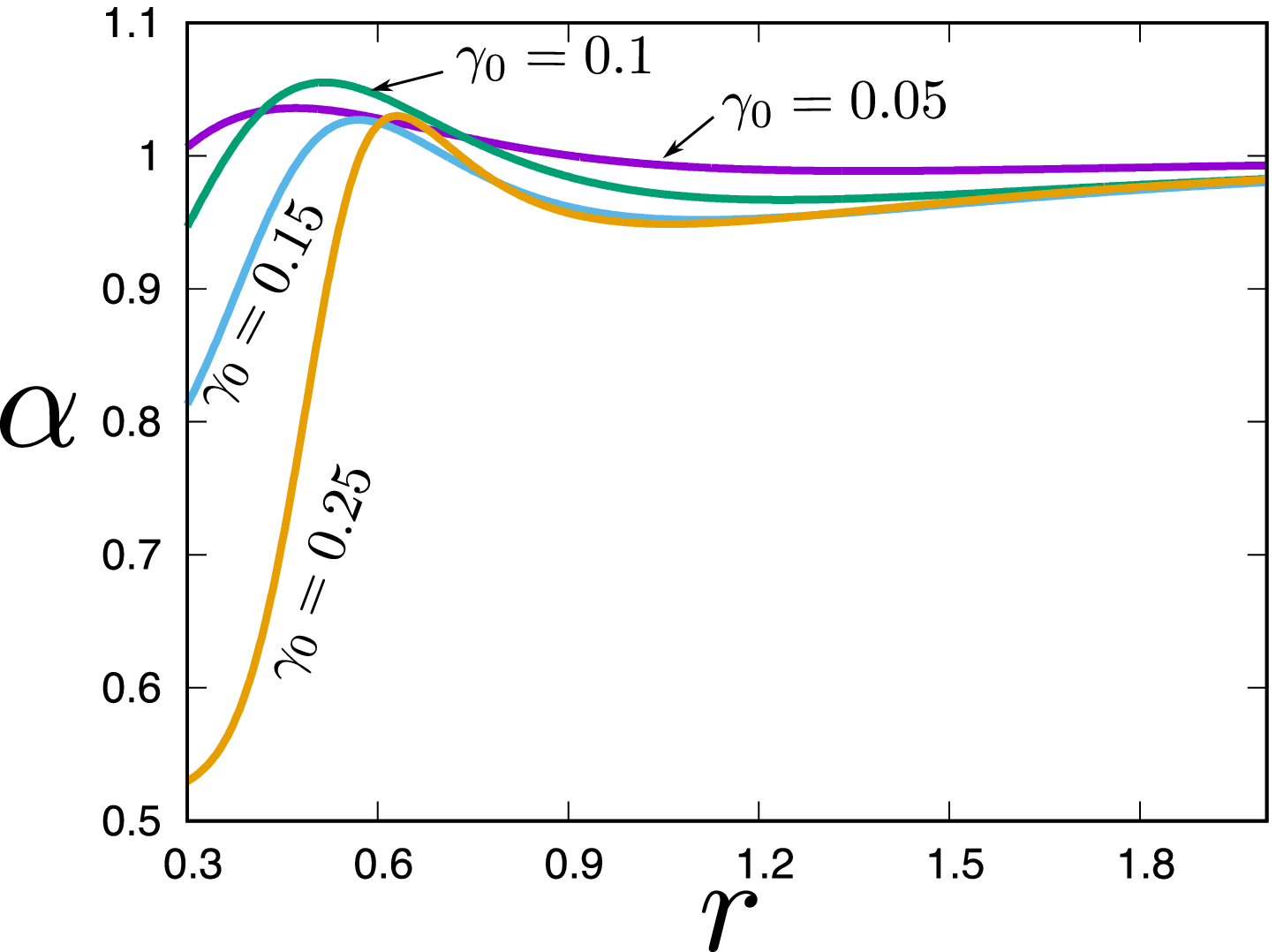}
  }
  \subfigure
 {\includegraphics[scale=0.24]{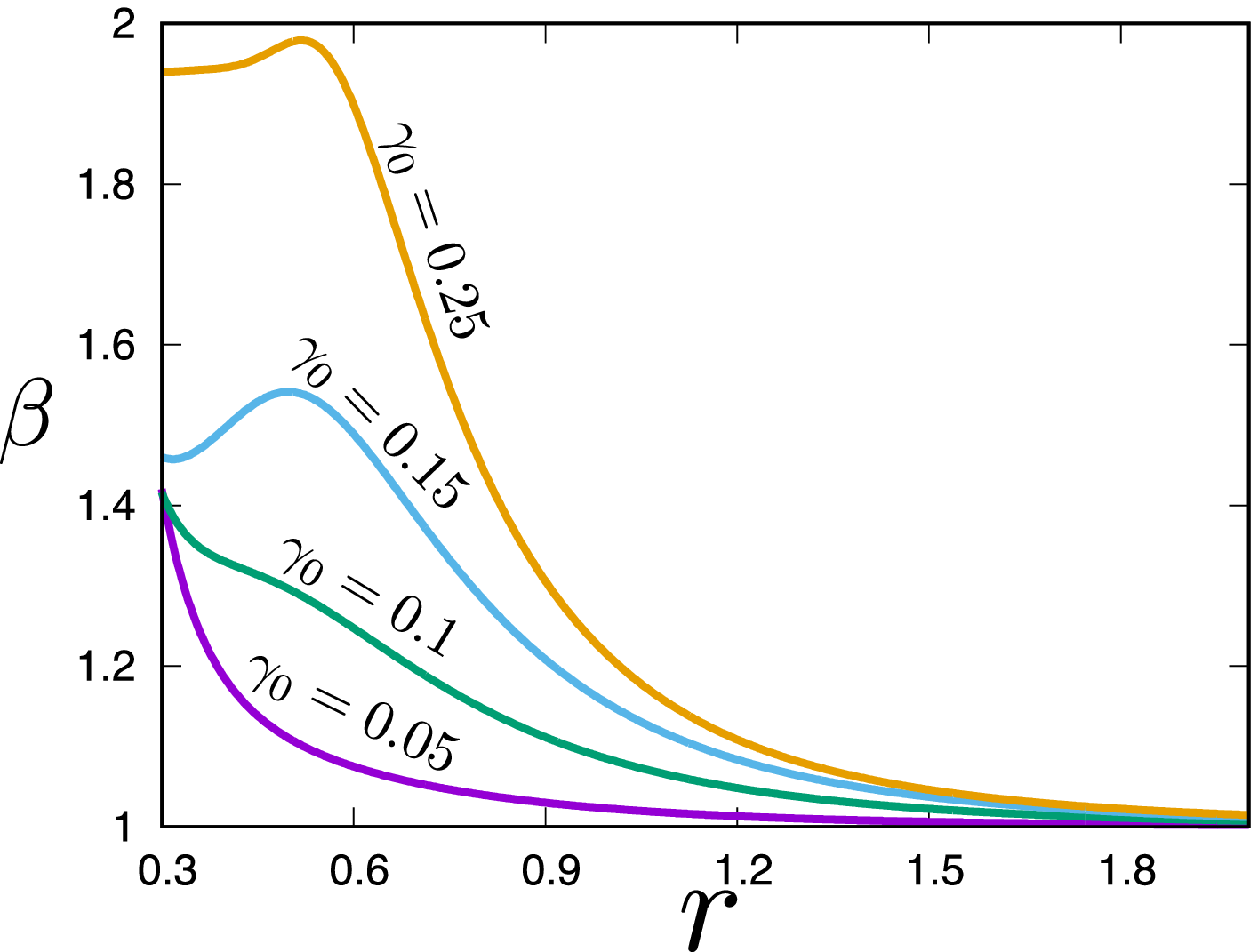}
  }
  \subfigure
 {\includegraphics[scale=0.24]{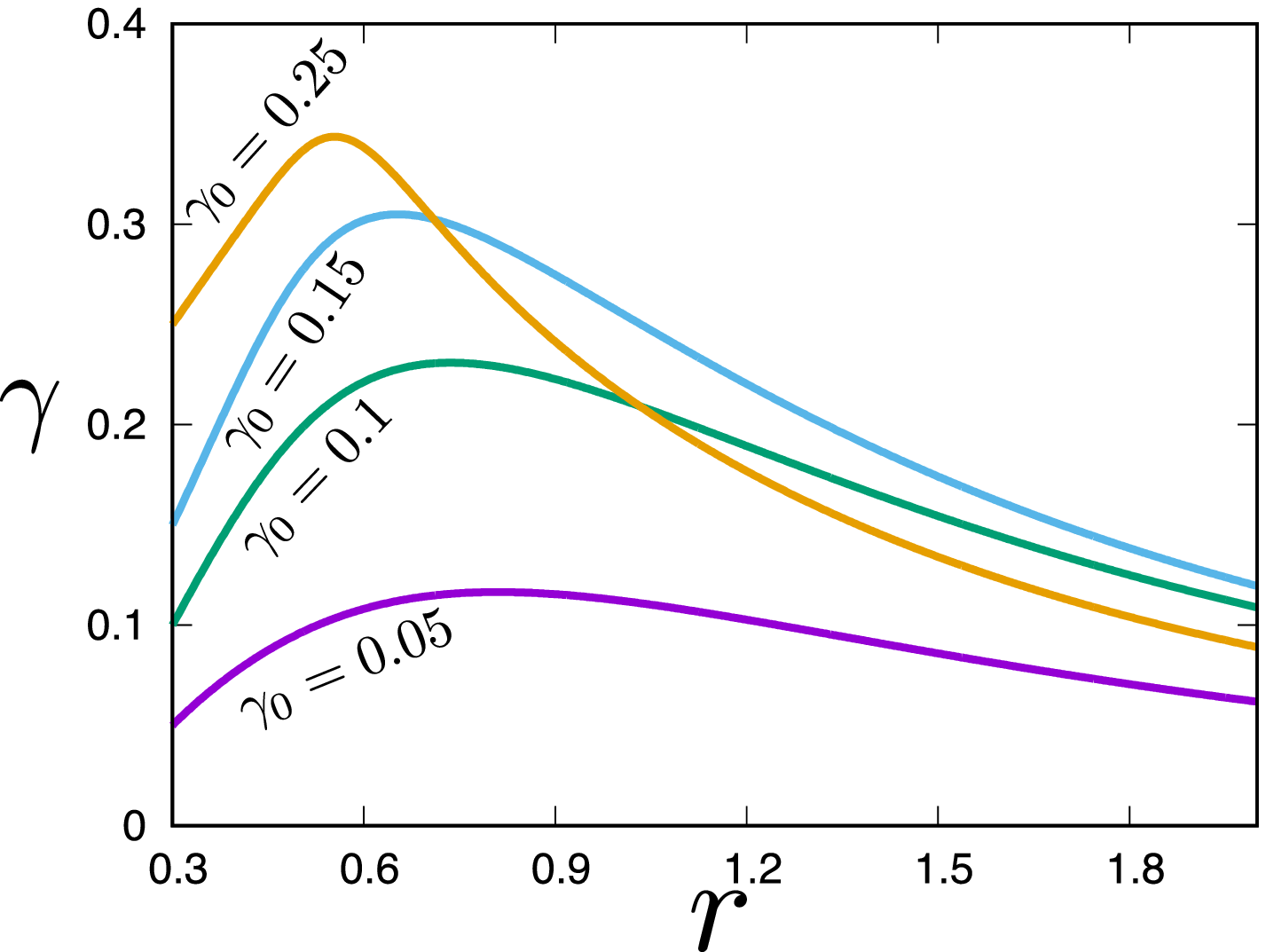}
  }
\subfigure
 {\includegraphics[scale=0.24]{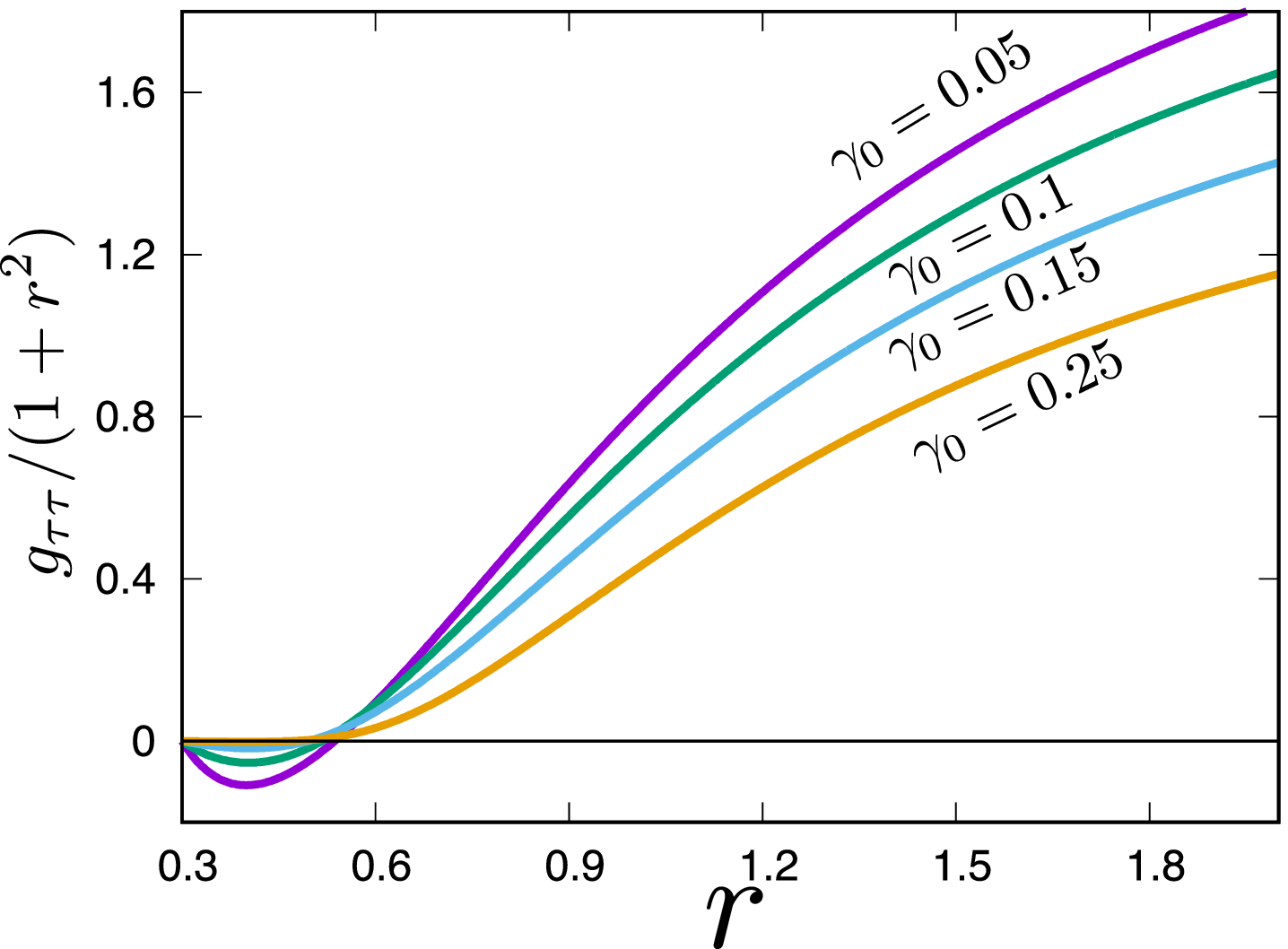}
  }
 \caption{%
Profile of $f,g,h,\alpha,\beta$ and $\gamma$ for photonic black resonators with $r_h=0.3$ and $\gamma_0=0.05,0.1,0.15,0.25$.
The last panel is the norm of the helical Killing vector, $K^2=g_{\tau\tau}$, normalized by $1+r^2$ for visibility. 
}
\label{photoresonatorfncs}
\end{figure}

In Fig.~\ref{photoresonatorfncs}, we show the profile of the functions $f,g,h,\alpha,\beta$ and $\gamma$ for $r_h=0.3$ and $\gamma_0=0.05,0.1,0.15,0.25$.
We also show the norm of the helical Killing vector $K^2=g_{\tau\tau}$ in the last panel. 
Although the Killing vector is timelike near the horizon, it becomes spacelike near the infinity.
This implies that the geometry of the photonic black resonator is dynamical.

\begin{figure}
\centering
\includegraphics[scale=0.6]{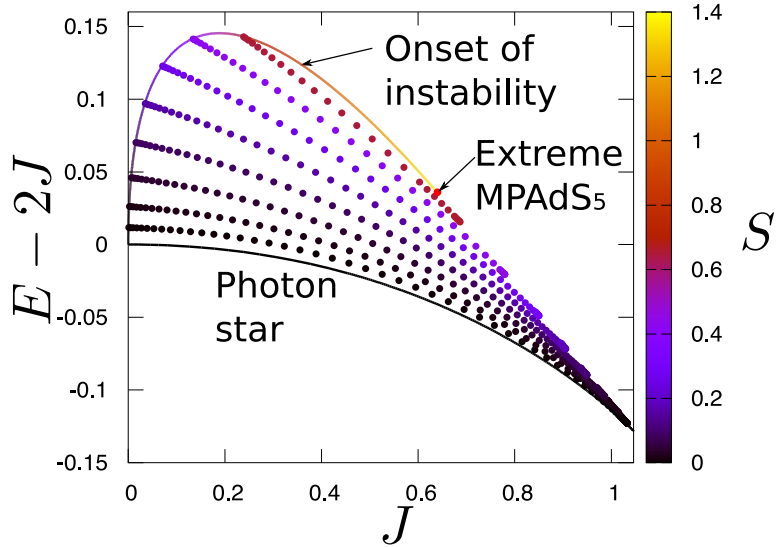}
 \caption{%
Phase diagram of photonic black resonators.
Dots correspond to the data points we numerically calculated.
The entropy $S$ is shown by the color map.
The black curve in the bottom is the family of photon stars.
The upper curve is MPAdS$_5$ at the onset of the Maxwell superradiant instability. 
The endpoint of the curve shown by the red dot corresponds to an extreme black hole.
}
\label{S2d}
\end{figure}

The phase diagram is shown in Fig.~\ref{S2d}.
The data points we numerically calculated are plotted in the dots, and the entropy $S$ is shown by the color.
For visibility, we take $E-2J$ as the vertical axis.
The black curve forming the lower boundary of the data points corresponds to the family of photon stars, which have zero entropy.
Meanwhile, the MPAdS$_5$ solutions at the onset of the Maxwell superradiant instability is also shown by the curve on the upper boundary.
This curve ends when it meets an extreme MPAdS$_5$ as in Fig.~\ref{onset}.

\begin{figure}
  \centering
  \subfigure[Fixed $J$]
 {\includegraphics[scale=0.45]{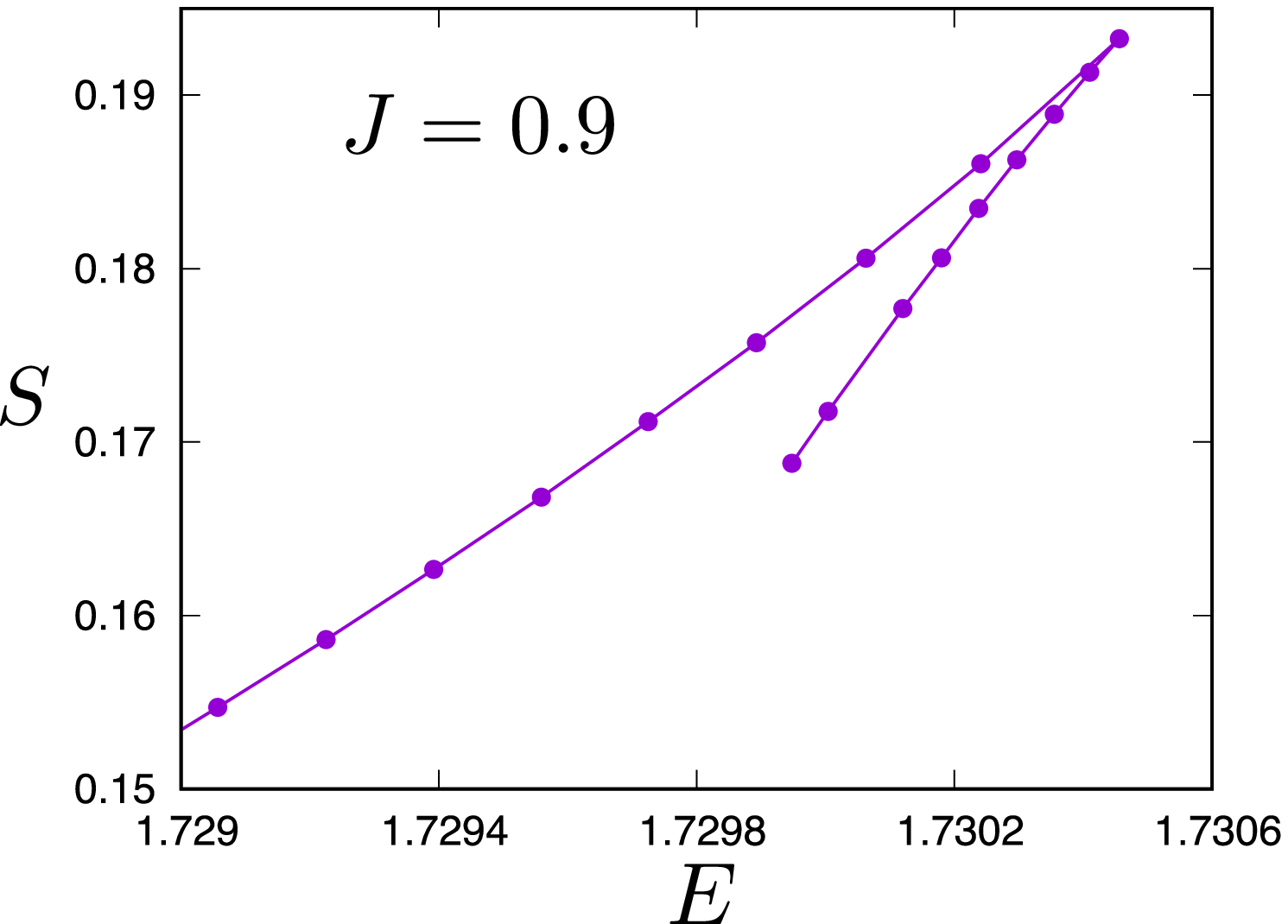}\label{S_Jfix}
  }
  \subfigure[Fixed $E$]
 {\includegraphics[scale=0.45]{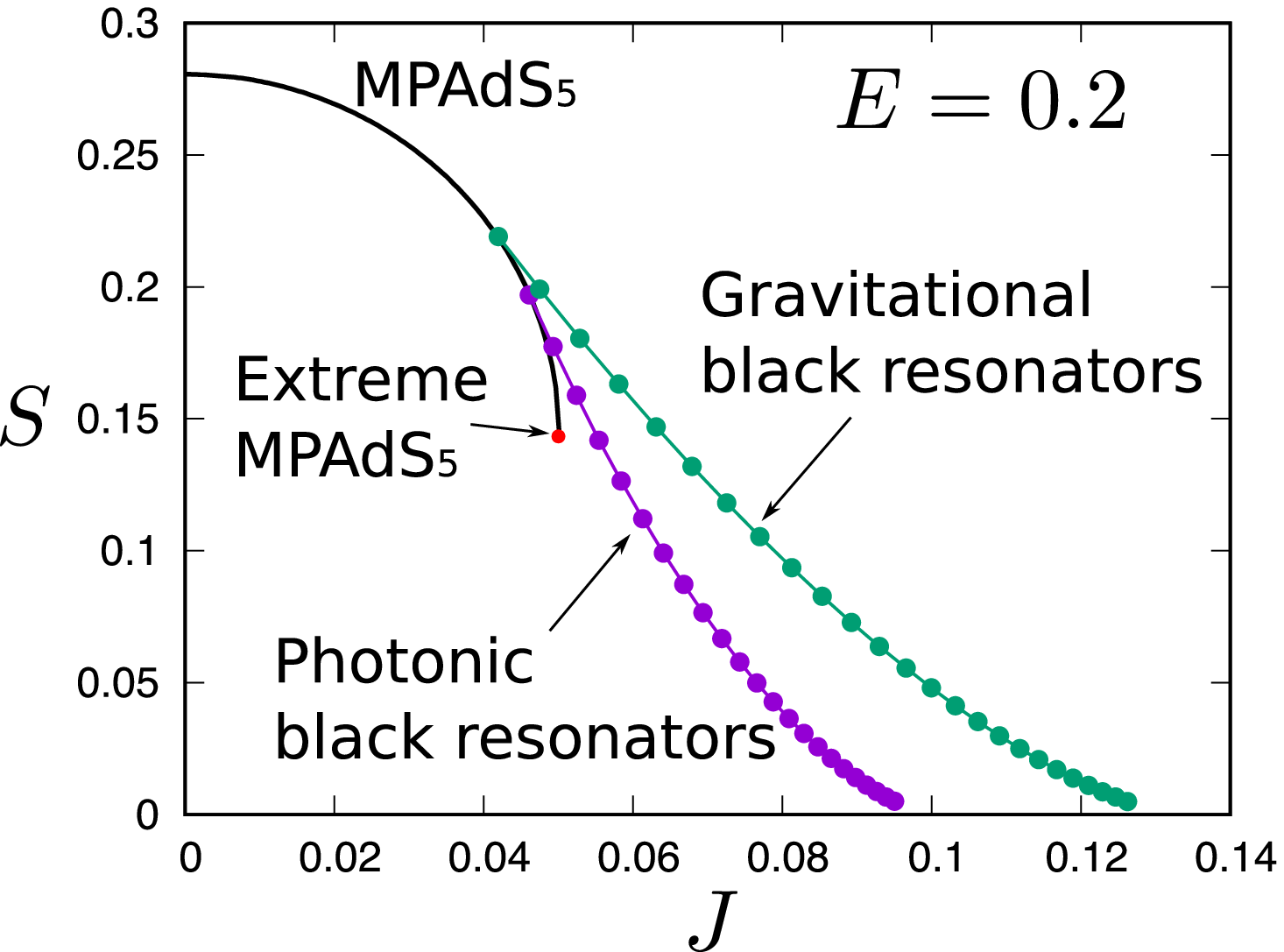}\label{S_Efix}
  }
 \caption{%
Entropy $S$ of the photonic black resonators for fixed (a) $J=0.9$ and (b) $E=0.2$.
In the right panel, the entropies of MPAdS$_5$ and gravitational black resonators are also shown.
}
\label{S_EJfix}
\end{figure}

The multivalued behavior of the physical quantities, discussed in Fig.~\ref{photonstarJE} for the photon stars, also appears in the photonic black resonators.
In the phase diagram, it is seen in the upper-right end of the plotted region.
To see this behavior in detail, in Fig.~\ref{S_Jfix} we show the entropy $S$ as the function of the mass $E$ at fixed $J=0.9$.
In this figure, the turning point can be found at $E\simeq 1.7304$. 

Finally, in Fig.~\ref{S_Efix} we compare the entropies between different solutions as a function of $J$ for fixed $E=0.2$.
In this figure, as well as the photonic black resonator, we show the entropies of MPAdS$_5$ and the $n=0$ purely gravitational black resonators studied in Ref.~\cite{Ishii:2018oms}. 
Around $J=0.042$ and $J=0.046$, the gravitational and photonic black resonators branch from MPAdS$_5$, respectively.
A photonic black resonator has a higher entropy than the corresponding MPAdS$_5$ at the same $(E,J)$.
We can also see that the photonic black resonators exist in the region of $(E,J)$ where no regular MPAdS$_5$ do.
Therefore, we find that the photonic black resonators extend the phase diagram from the case only MPAdS$_5$ are included (otherwise there are singular over-rotating MPAdS$_5$).
However, we also find that the gravitational black resonators have a higher entropy than the photonic black resonators at the same $(E,J)$.
This implies that the photonic black resonators would be further unstable against $SU(2)$-symmetric perturbations which suppress the Maxwell field, and they would evolve into the gravitational black resonators if dynamical time evolution is considered.

\begin{figure}
  \centering
  \subfigure[Entropy]
 {\includegraphics[scale=0.4]{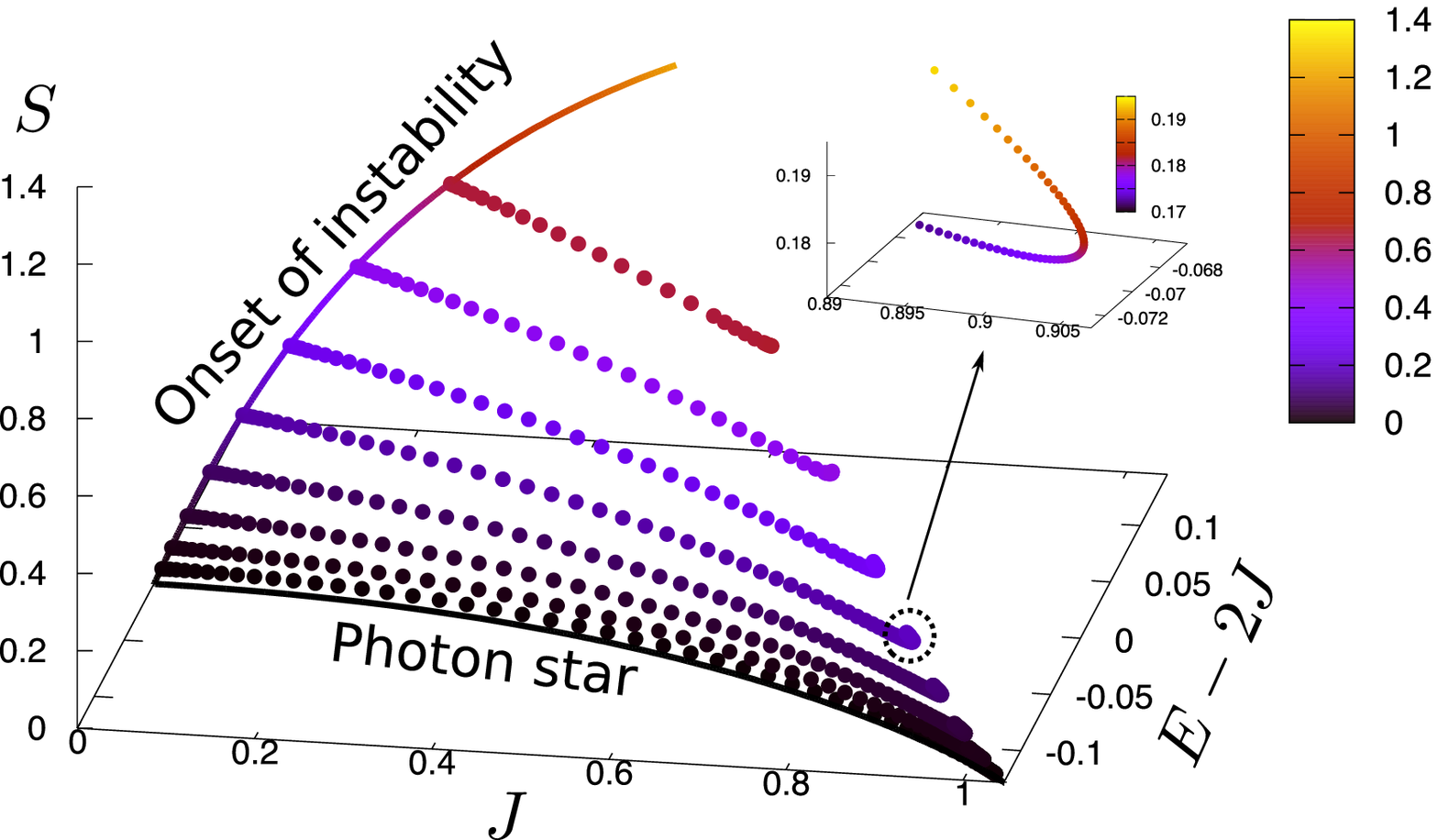}\label{S3d}
  }
  \subfigure[Angular velocity]
 {\includegraphics[scale=0.4]{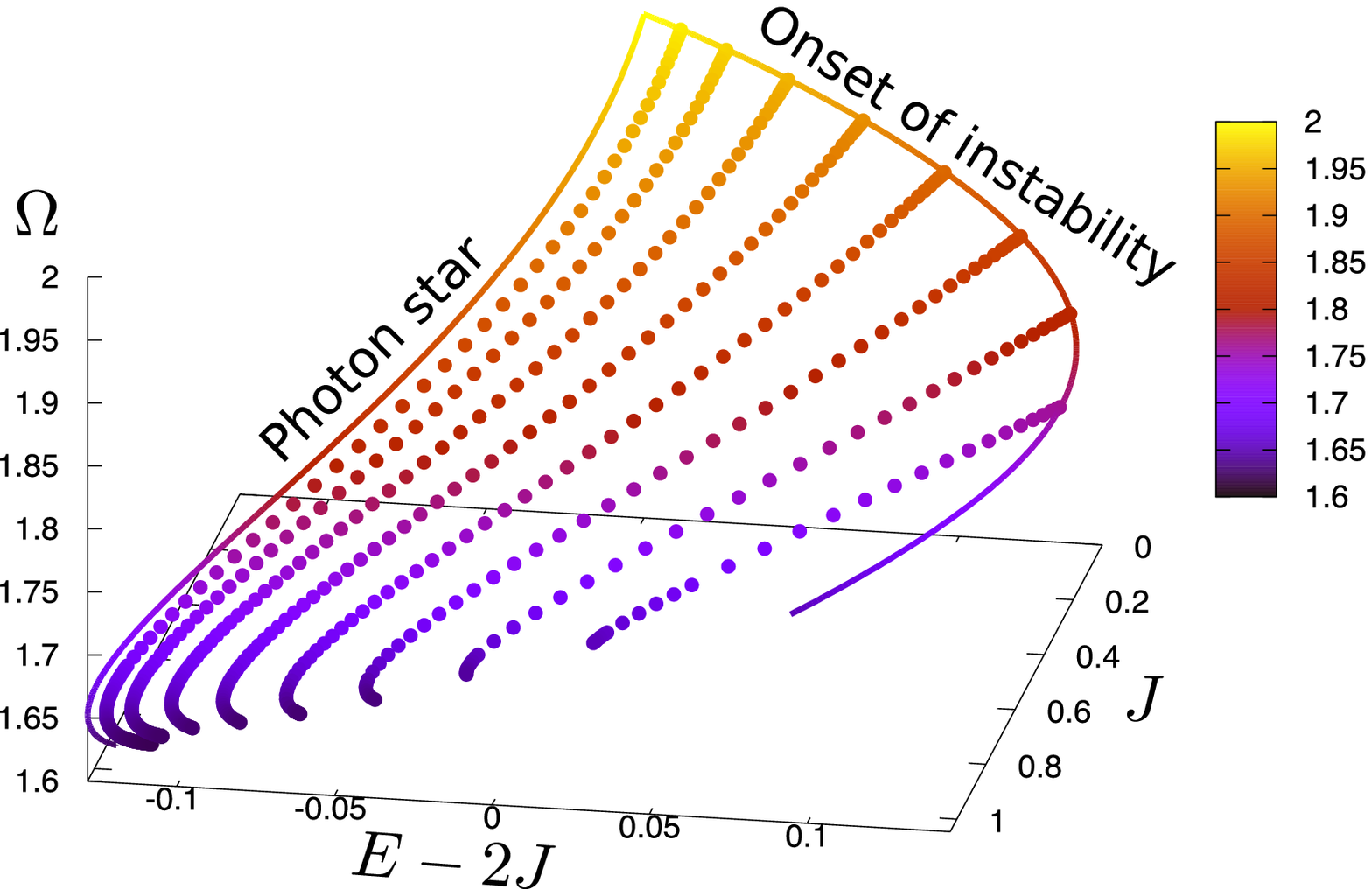}
  }
  \subfigure[Temperature]
 {\includegraphics[scale=0.4]{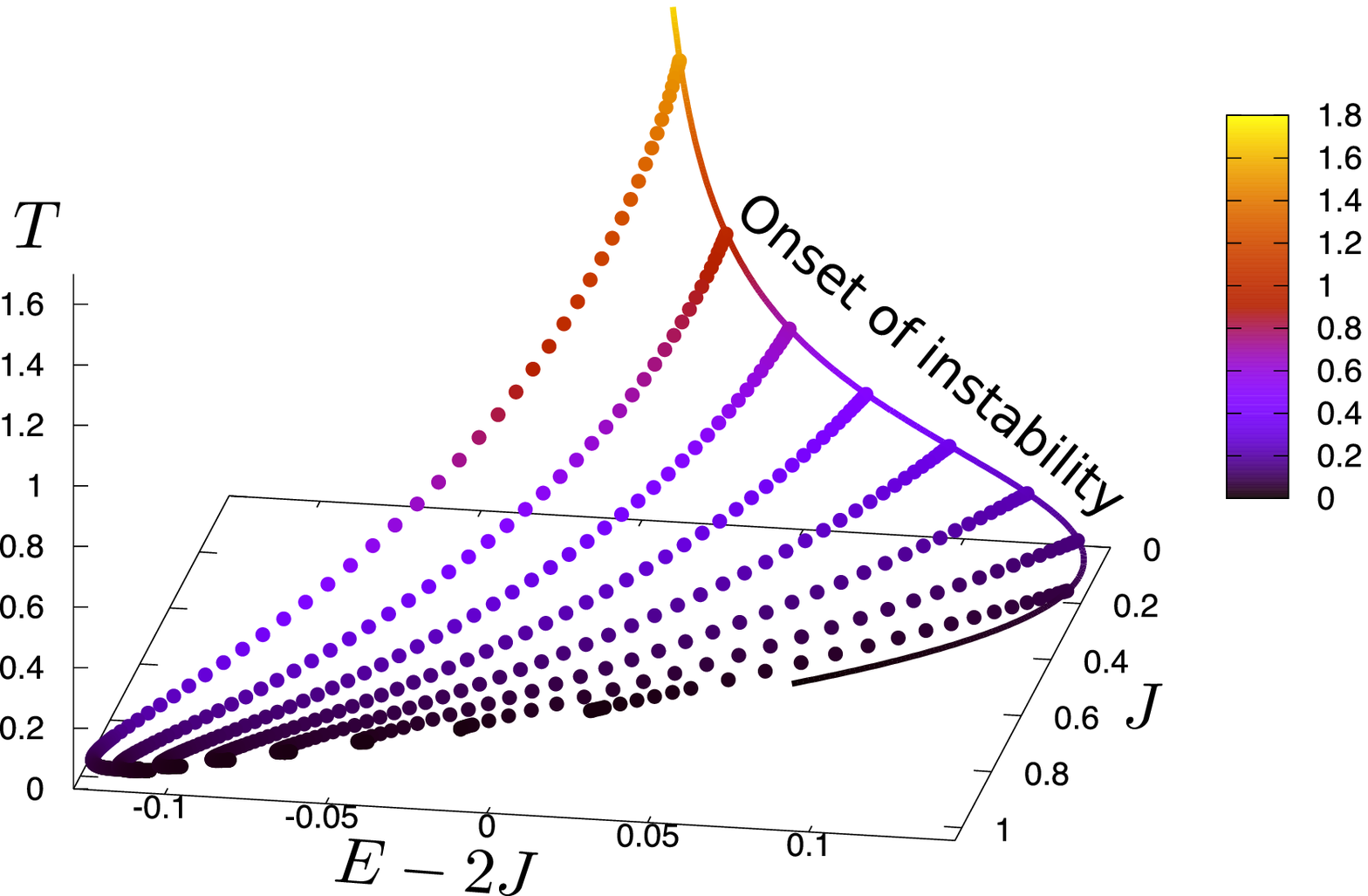}\label{T3d}
  }
  \subfigure[Current]
 {\includegraphics[scale=0.4]{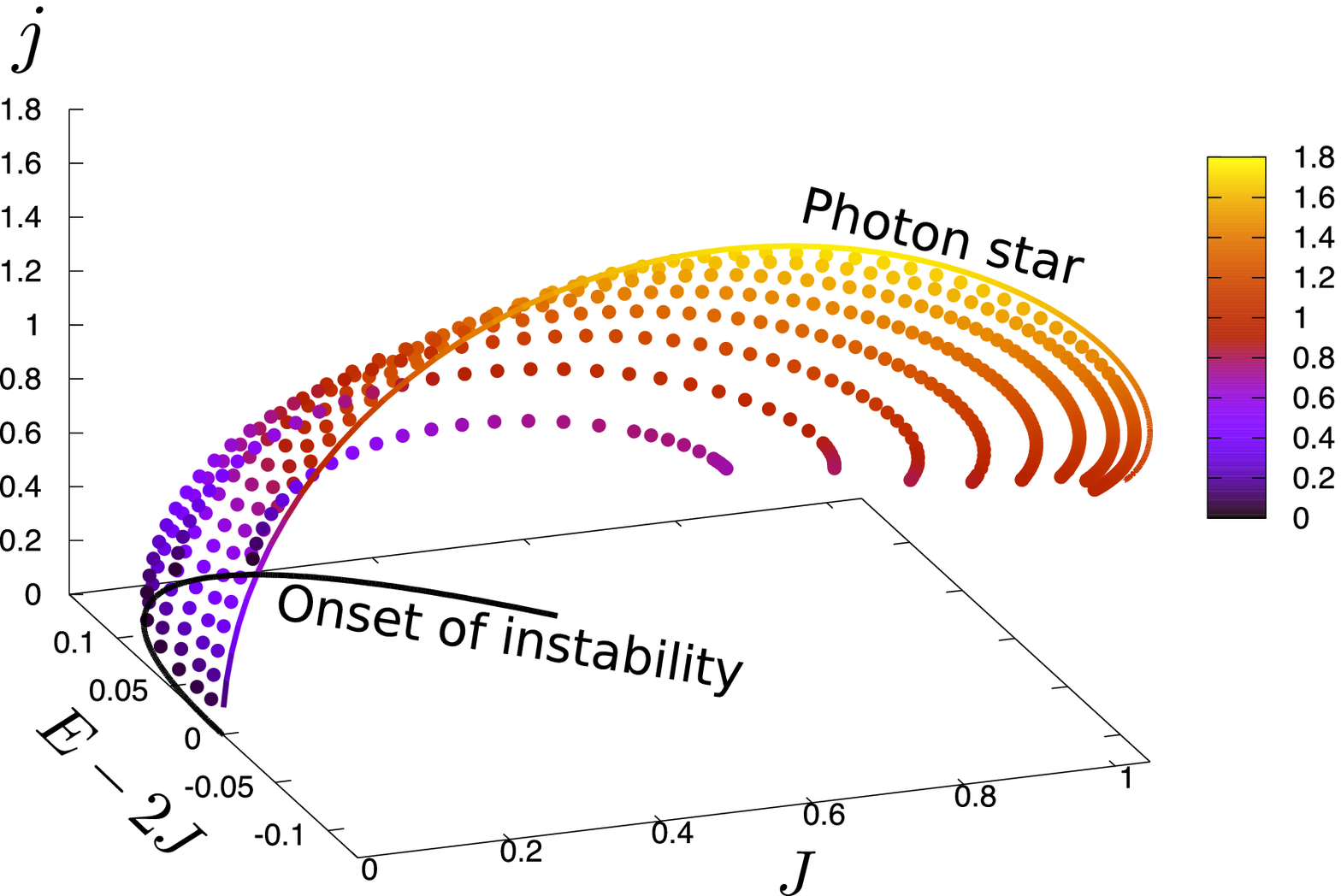}
  }
 \caption{%
Thermodynamical quantities of photonic black resonators.
}
\label{S_etc_3d}
\end{figure}

Three-dimensional plots of the entropy $S$, angular velocity $\Omega$, temperature $T$, and electric current $j$ of photonic black resonators 
are shown in Fig.~\ref{S_etc_3d}. 
In Fig.~\ref{T3d}, we see that the temperature approaches zero at a boundary of the region where photonic black resonators exist.
This suggests that 
there are limiting photonic black resonators whose temperature approaches zero.
However, the perturbative analysis of the extreme MPAdS$_5$ in appendix~\ref{app:ext} indicates that
the limiting solution would be singular.

\section{Conclusions}
\label{conc}

We constructed a family of photonic black resonators which bifurcate from MPAdS$_5$ at the onset of the Maxwell superradiant instability.
We also obtained photon stars as the horizonless limit of such photonic black resonators.
These solutions have the isometry group $R \times SU(2)$ and are based on the cohomogeneity-1 metric ansatz together with a Maxwell field consistent with the symmetries of the metric.
The Einstein and Maxwell equations then reduced to ODEs.
Solving the ODEs and evaluating thermodynamical quantities, we determined the phase structure of the photonic black resonators and photon stars.
The phase diagram is shown in Fig.~\ref{S2d}.
The photonic black resonators exist only in a finite domain in the $(E,J)$ diagram, and there is an upper bound on $J$ unlike the case of the purely gravitational black resonators studied in Ref.~\cite{Ishii:2018oms}.
Because of this, physical quantities become multivalued near a boundary of the region where photonic black resonators exist.
The angular velocity of the solutions we obtained always satisfies $\Omega>1$, and this implies that the photonic black resonators and photon stars describe dynamical spacetime.
It has been already indicated by the theorem of Ref.~\cite{Green:2015kur} that such dynamical spacetime must be unstable.
In the case of the photonic black resonators, we observed that they should be also unstable against $SU(2)$-symmetric perturbations by comparing the entropies of a photonic black resonator and a gravitational black resonator with the same $(E,J)$: The former has a smaller entropy than the latter.

This observation would open up the time evolution of the Maxwell superradiant instability as an important future direction.
In asymptotically AdS$_4$ spacetime, the nonlinear time evolution of the superradiant instability has been studied in Ref.~\cite{Chesler:2018txn}.
It has been argued that superradiant instabilities bring black resonators to evolve into small scales, and it makes the study of their time evolution difficult \cite{Niehoff:2015oga}.
In the case of AdS$_5$, we can exactly impose the $SU(2)$-symmetry in the dynamics, and we will just need to solve the time evolution of $(1+1)$-dimensional PDEs to see the dynamics of black resonators.
Because a photonic black resonator has a lower entropy than the gravitational black resonator with the same conserved charges $(E,J)$, we will see the time evolution that 
the Maxwell field would be absorbed by the black hole, while it remains neutral, and the spacetime will approach a purely gravitational black resonator. 
Similarly, we will also be able to study the weakly turbulent instability of AdS~\cite{Bizon:2011gg} triggered by the Maxwell field perturbation by making use of the setup in this paper.

Time periodic Maxwell fields in asymptotically AdS spacetime will be important in applied AdS/CFT---the application of the AdS/CFT correspondence to realistic systems. For instance, to understand nonequilibrium dynamics of strongly correlated quantum many-body systems under a periodic driving is one of the most significant problems in condensed matter physics. This has been addressed by using the AdS/CFT correspondence~\cite{Li:2013fhw,Natsuume:2013lfa,Hashimoto:2016ize,Kinoshita:2017uch,Ishii:2018ucz,Auzzi:2012ca,Auzzi:2013pca,Rangamani:2015sha,Biasi:2019eap}.
Our methodology to treat the time periodic Maxwell field in asymptotically AdS spacetime 
may provide a new hint to investigate nonequilibrium processes in strongly correlated systems.
For example, in Refs.~\cite{Li:2013fhw,Natsuume:2013lfa,Ishii:2018ucz}, 
a time periodic electric field was considered in the holographic superconductor~\cite{Gubser:2008px,Hartnoll:2008vx,Hartnoll:2008kx} in the probe limit. 
The application of our method to holographic superconductors may help us to 
investigate new phases of matter in the presence of a time-periodic electromagnetic field.

\acknowledgments
The authors would like to thank Benson Way for useful discussions and comments.
The work of T.~I.~was supported by the Supporting Program for Interaction-based Initiative Team Studies (SPIRITS) from Kyoto University, and JSPS KAKENHI Grant Number JP18H01214 and JP19K03871.

\appendix

\section{Photonic black resonators and photon stars with overtones}
\label{overtone}

In section~\ref{Maxwell}, we studied the Maxwell perturbation in the MPAdS$_5$ background and 
found the onsets of the superradiant instability for the fundamental tone $n=0$ and overtones $n=1,2,\cdots$.
In the main text of this paper, we focused on the photon stars and photonic black resonators for the fundamental tone.
In this appendix, we summarize the results for the first overtone $n=1$.

\begin{figure}
  \centering
  \subfigure[Mass]
 {\includegraphics[scale=0.33]{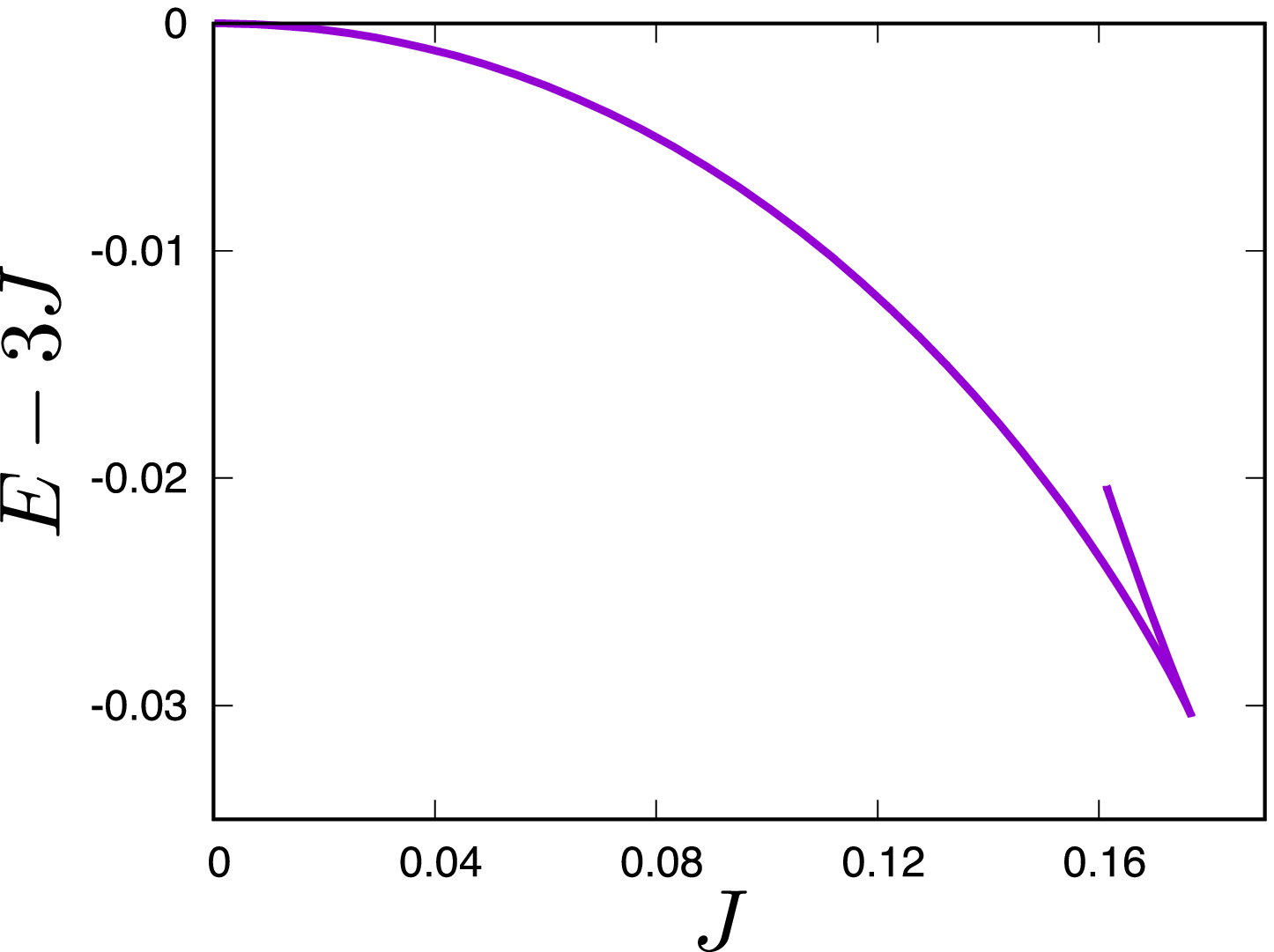}
  }
  \subfigure[Angular velocity]
 {\includegraphics[scale=0.33]{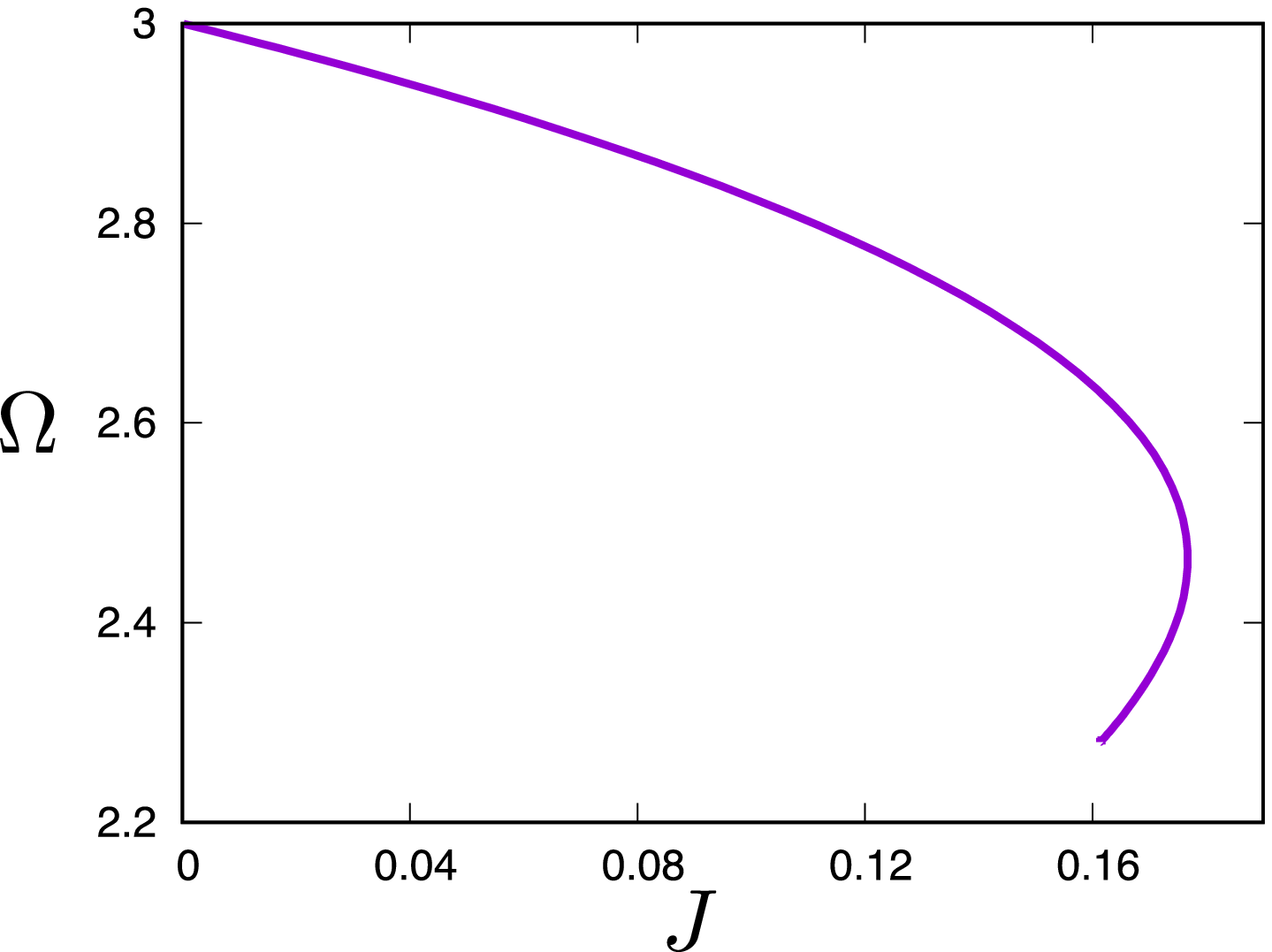}
  }
  \subfigure[Current]
 {\includegraphics[scale=0.33]{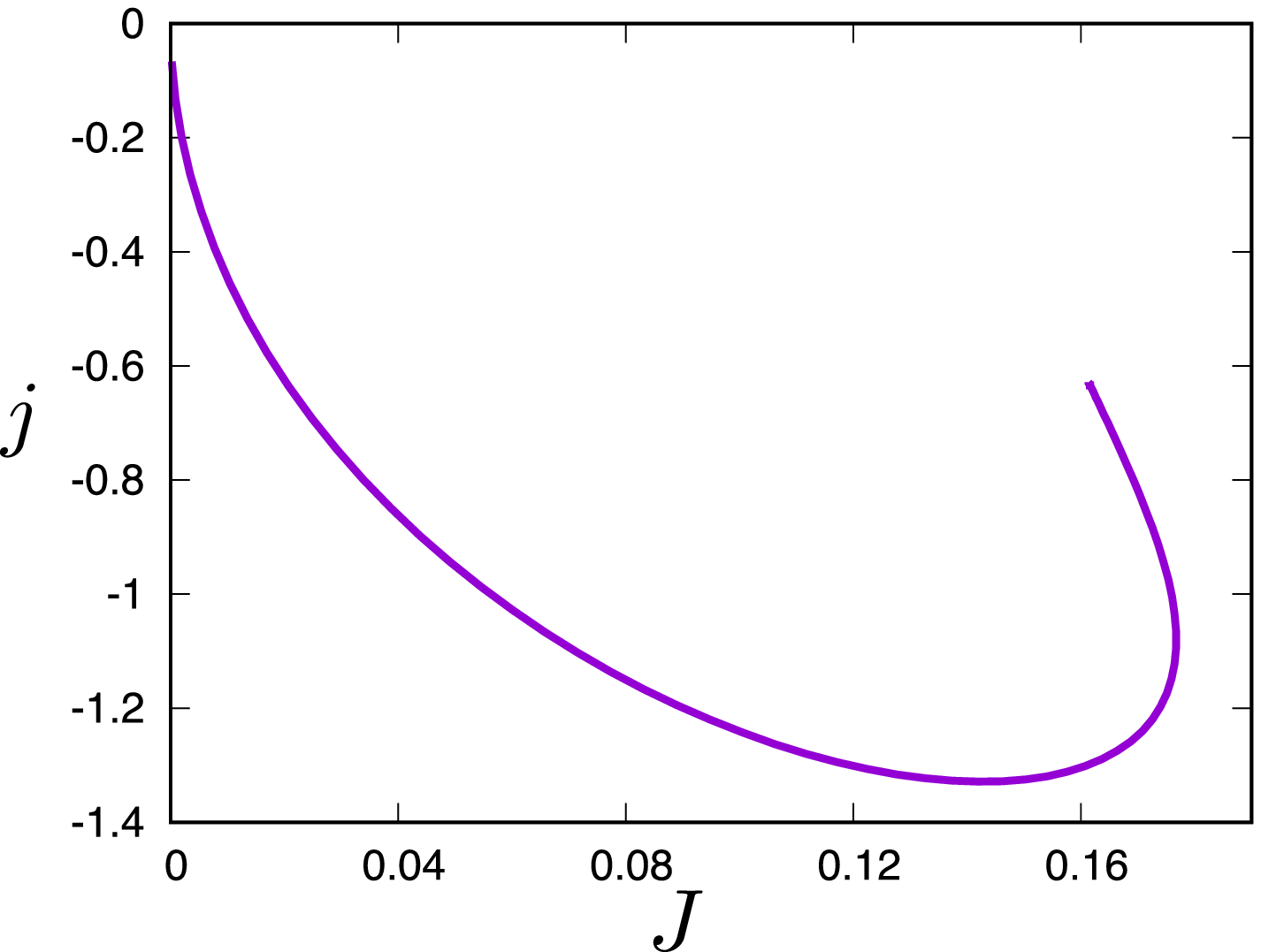}
  }
 \caption{%
The mass $E$, angular velocity $\Omega$ and electric current $j$ are shown 
as a function of the angular momentum $J$ for the photon stars with the first overtone $n=1$.
For visibility, $E-3J$ is taken in the vertical axis in the plot of $E$.
}
\label{photonstarJEn1}
\end{figure}

Shown in Fig.~\ref{photonstarJEn1} are the mass $E$, angular velocity $\Omega$, and electric current $j$ as a function of the angular momentum $J$ for the photon stars with $n=1$.
The angular velocity approaches $\Omega\to 3$ in the pure AdS limit $J\to 0$, which is a normal mode frequency of the pure AdS. In the first panel in the figure, we took $E-3J$ as the vertical axis for visibility.
We again find a turning point for $J$ in the diagrams as is the case for $n=0$.
The angular velocity always satisfies $\Omega>1$, too, implying that the $n=1$ photon star is dynamical.
The sign of $j$ here is opposite to that of the fundamental tone because of the number of nodes in $\gamma(r)$.

\begin{figure}
\begin{center}
\includegraphics[scale=0.6]{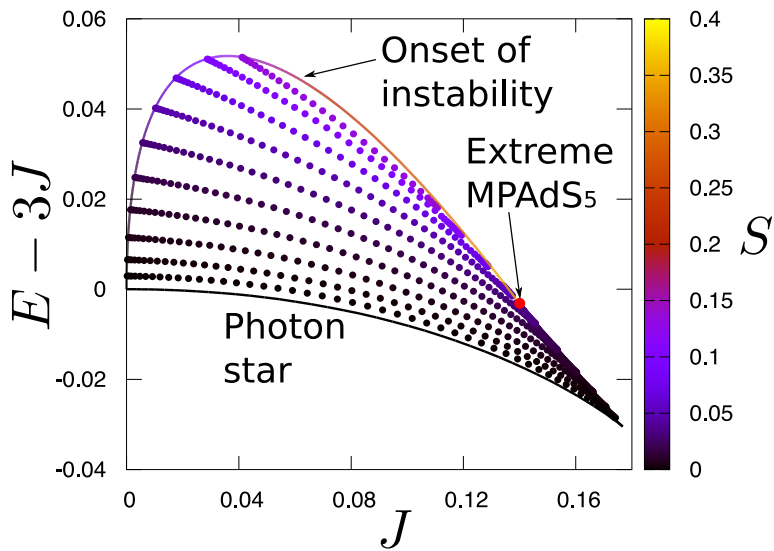}
\end{center}
 \caption{%
Phase diagram of the photonic resonators with overtone number $n=1$.
Dots are data points for $(E,J)$.
Their color corresponds to the value of the entropy $S$.
The black curve in the bottom represents the family of $n=1$ photon stars.
The upper curve corresponds to the MPAdS$_5$ at the onset of the $n=1$ Maxwell superradiant instability. 
The endpoint of this curve shown by the red point is an extreme black hole.
}
\label{S2d_n1}
\end{figure}

Fig.~\ref{S2d_n1} is the phase diagram of photonic resonators with $n=1$.
The values of $(E,J)$ of the photonic black resonators are located by dots at the data points we obtained, and the color of the dots corresponds to the value of entropy $S$. We take $E-3J$ as the vertical axis for visibility.
The photonic black resonators with $n=1$ also exist only in a finite domain in the diagram.
The physical quantities become multivalued at the upper-right boundary of the plot region.

\begin{figure}
  \centering
  \subfigure[Entropy]
 {\includegraphics[scale=0.4]{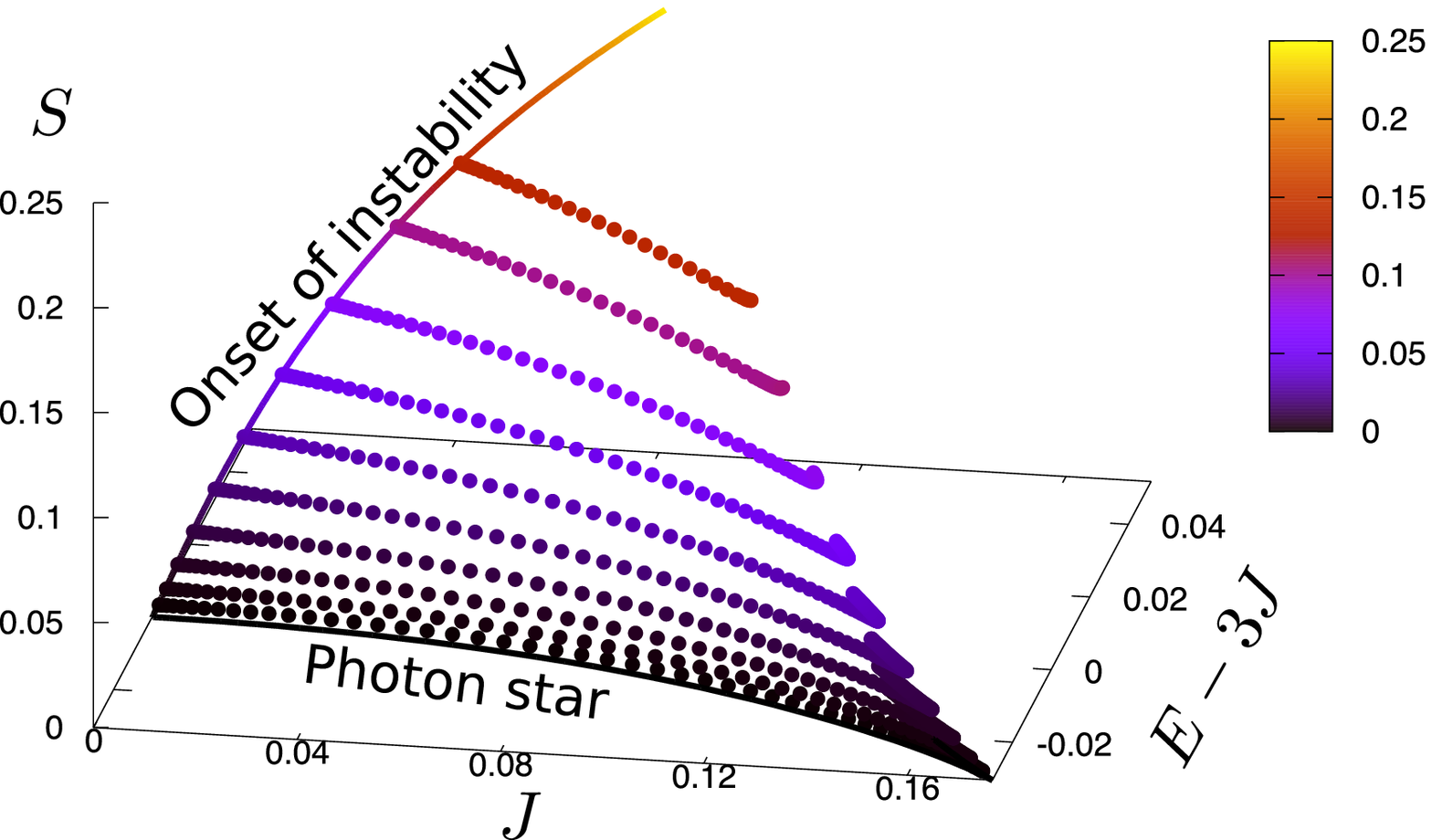}
  }
  \subfigure[Angular velocity]
 {\includegraphics[scale=0.4]{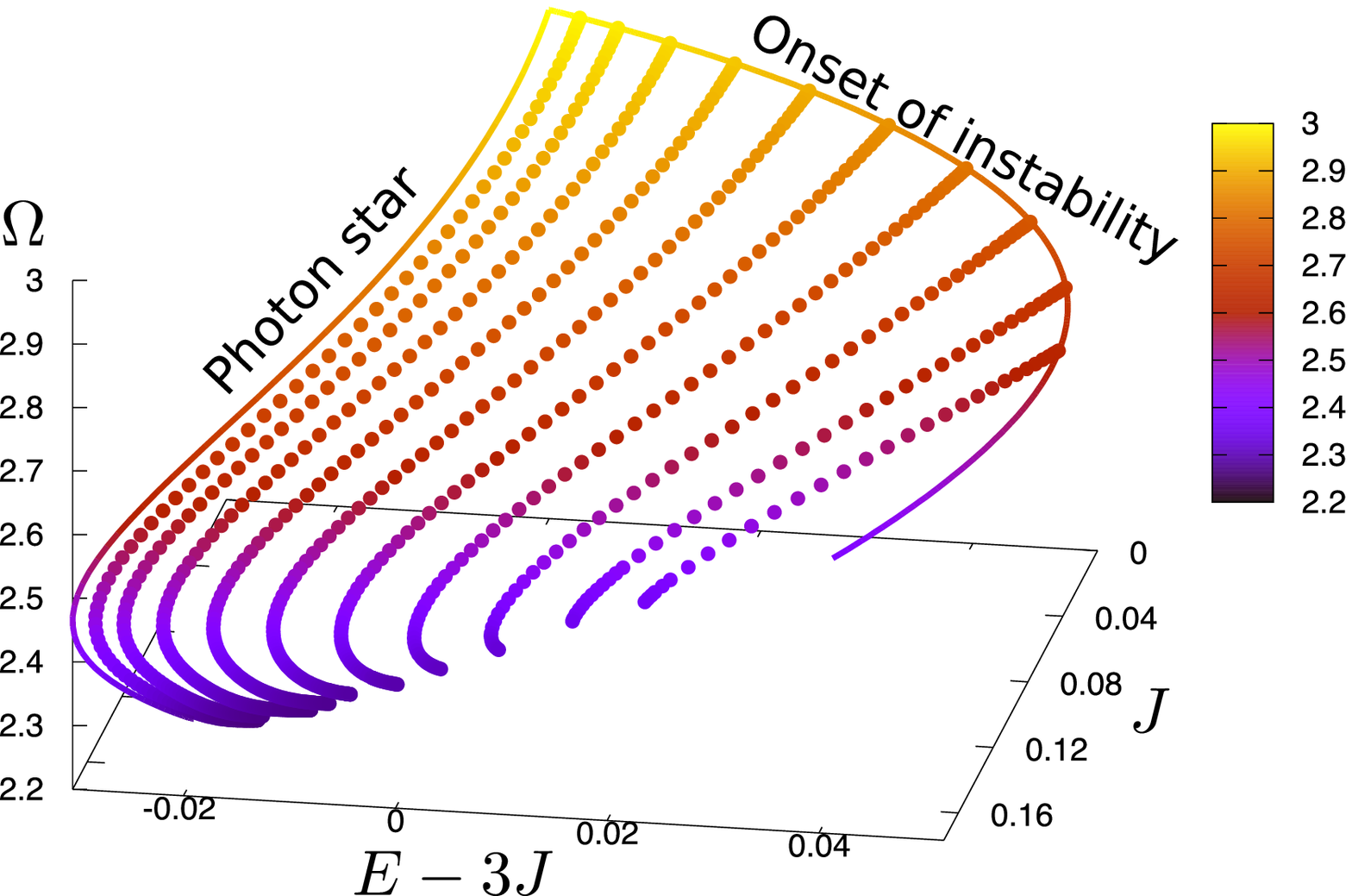}
  }
  \subfigure[Temperature]
 {\includegraphics[scale=0.4]{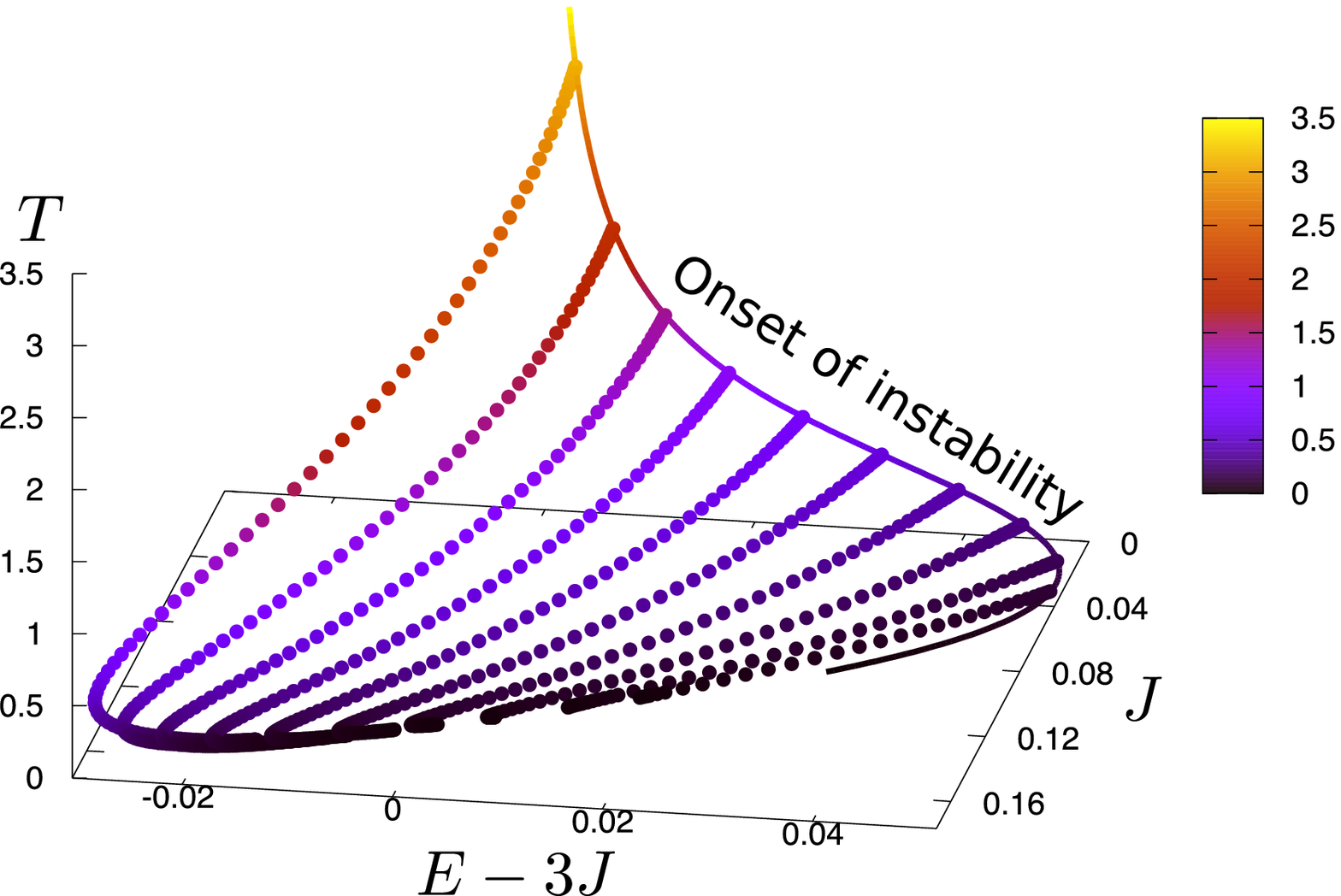}
  }
  \subfigure[Current]
 {\includegraphics[scale=0.4]{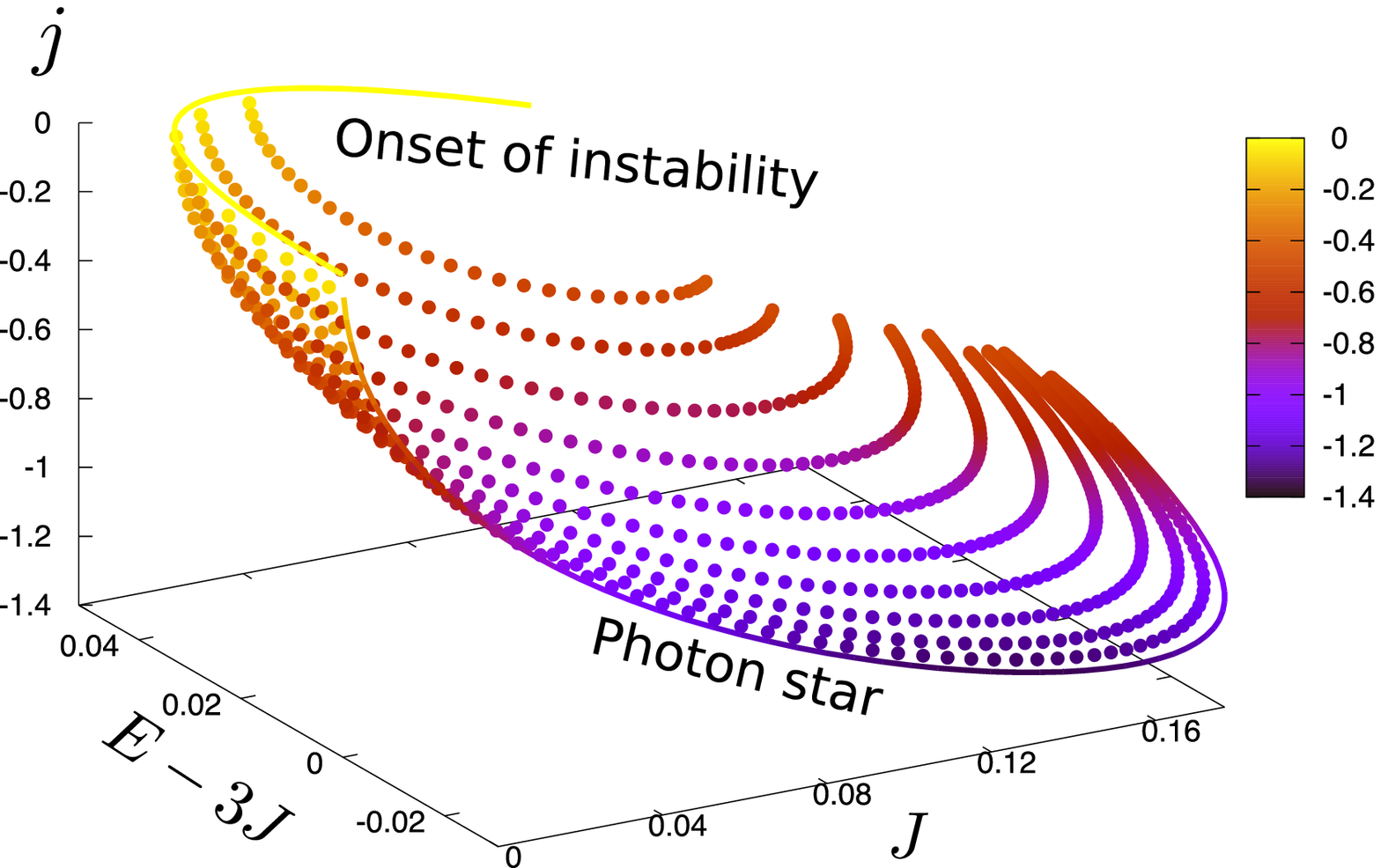}
  }
 \caption{%
Thermodynamical quantities of $n=1$ photonic black resonators.
}
\label{S_etc_3d_n1}
\end{figure}

Fig.~\ref{S_etc_3d_n1} contains three-dimensional plots of the entropy $S$, angular velocity $\Omega$, temperature $T$ and electric current $j$ of the photonic black resonators with $n=1$.
The angular velocity always satisfies $\Omega>1$. This implies that the photonic black resonators with $n=1$ are dynamical.
In Fig.~\ref{T3d}, the temperature approaches zero in one of the boundaries of the data region.
This indicates that there may be a zero temperature limit for the photonic black resonator solution 
although that may be singular.

\section{Perturbation of extreme MPAdS$_5$}
\label{app:ext}
In this appendix, we consider the perturbation on an extreme MPAdS$_5$ background in order to examine if the onset of the superradiant instability meets the extreme MPAdS$_5$.
The condition for the extreme MPAdS$_5$ reads 
\begin{equation}
\Omega_\mathrm{ext} = \sqrt{1+\frac{1}{2 r_h^2}}. 
\end{equation}
Using this, we take the extreme limit in the linear perturbation equation \eqref{pbr_onset_eq}.
To identify the boundary condition at the extreme horizon, we solve the resulting equation near the extreme horizon.
We find that the field behaves $\gamma \sim (r-r_h)^{n_\pm}$ with $n_\pm = \left(-1 \pm \sqrt{1-2/(1+3r_h^2)^{2}} \right)/2$.
The exponent is real if $r_h>\sqrt{(\sqrt{2}-1)/3}\sim 0.3716$.
In that region, however, both exponents are negative, $n_\pm<0$ (with $n_+ \to 0$ and $n_- \to -1$ as $r_h \to \infty$).
This means that $\gamma$ is singular at the horizon.
It is then expected that full solutions with nontrivial $\gamma$ on the extreme MPAdS$_5$ might also be singular.
Nevertheless, if we factor out the singular part as $\gamma = (r-r_h)^{n_+} \tilde{\gamma}$ so that $\tilde{\gamma}$ is regular while we simply suppress the other solution with $n_-$ for simplicity, we can show the presence of normal modes 
which would correspond to the extreme limit of the onset of the superradiant instability.
Imposing $\tilde{\gamma}=1$ at $r=r_h$ and $\tilde{\gamma}=0$ at $r=\infty$, we find normal modes at $r_h = 0.5559$ and $0.3720$.
These values are consistent with the endpoints for the $n=0$ and $n=1$ curves as they approach the extreme BH limit.
For $n\ge2$ modes, we were not able to pin down where the onset curves end up on the extreme BH limit because these modes approach the extreme BH limit very slowly where numerical calculations become difficult.

\bibliography{bunken_PBR}

\end{document}